\documentclass[10pt]{article}
\usepackage{xparse}
\ExplSyntaxOn
\NewDocumentCommand{\mref}{m}{\quinn_mref:n {#1}}
\seq_new:N \l_quinn_mref_seq
\cs_new:Npn \quinn_mref:n #1
 {
  \seq_set_split:Nnn \l_quinn_mref_seq { , } { #1 }
  \seq_pop_right:NN \l_quinn_mref_seq \l_tmpa_tl
  ( 
  \seq_map_inline:Nn \l_quinn_mref_seq
    { \ref{##1},\nobreakspace } 
  \exp_args:NV \ref \l_tmpa_tl 
  ) 
 }
\ExplSyntaxOff
\usepackage[margin=1.5in]{geometry}
\usepackage[utf8]{inputenc}
\usepackage{authblk}
\usepackage{amsmath,bm}
\usepackage[title]{appendix}
\usepackage{float}
\usepackage{graphicx}
\usepackage{wrapfig}
\usepackage{xparse}
\usepackage{caption}
\usepackage{subcaption}

\usepackage{tabularx}
\usepackage[sorting=none, style=numeric-comp]{biblatex}
\usepackage{subcaption}
\usepackage{cleveref}
\usepackage{graphicx}
\bibliography{references}
\crefname{subtable}{panel}{panels}
\Crefname{subtable}{Panel}{Panels}
\title{Lattice Boltzmann Model in General Curvilinear Coordinates Applied to Exactly Solvable 2D Flow Problems}

\author{Alexei Chekhlov}
\author{Ilya Staroselsky}
\author{Raoyang Zhang}
\author{Hudong Chen}
\affil{Dassault Systemes, 175 Wyman Street, Waltham, MA 02451}

\date{December 2022}

\begin{document}

\maketitle

\begin{abstract}
Numerical simulation results of basic exactly solvable fluid flows using the previously proposed Lattice Boltzmann Method (LBM) formulated on a general curvilinear coordinate system \cite{HChen} are presented. As was noted in \cite{HChen}, such curvilinear Lattice Boltzmann Method preserves a fundamental one-to-one exact advection feature in producing minimal numerical diffusion, as the Cartesian lattice Boltzmann model.  As we numerically show, the new model converges to exact solutions of basic fluid flows with the increase of grid resolution in the presence of both natural curvilinear geometry and/or grid non-uniform contraction, both for near equilibrium and non-equilibrium LBM parameter choices. 

\end{abstract}

\section{Introduction}

Lattice Boltzmann Method (LBM) is currently one of the most accurate and widely used methods of simulation and analysis of continuum media. In many physics and engineering applications, LBM almost completely displaced and/or replaced its direct finite-difference or finite-volume Navier-Stokes-based competitors. As it is well-known, currently used versions of variable grid LBM are based on the volumetric formulation of LBM \cite{volumetric} with the Cartesian formulation of lattice cells and space basis vectors and linear approximations of curved boundaries within a lattice cell and Variable Resolution (or VR)-regions \cite{ChenFilippova,Molvig1}. Their accuracy and consistency could be potentially improved if a true non-Cartesian formulation were available. 

As was detailed in \cite{HChen}, some attempts to achieve this goal were made in \cite{He,Barraza}, albeit with a loss of a very important feature of the basic LBM - precision of the advection stage, and brought a significant amount of numerical dissipation. The adequate way to avoid this is to represent the process on a general coordinate system based on Riemann geometry \cite{book} where particles move on a curved path in Euclidean space. Such a method should be producing a curvilinear inertial body force. Several such attempts were made in a series of papers \cite{Mendosa1,Mendosa2,Mendosa3}. 

Recently, a new approach \cite{HChen} was proposed which for the first time has used a truly curvilinear tensor formulation of the Lattice Boltzmann Method. As noted in \cite{HChen}, the following key differences to works \cite{Mendosa1,Mendosa2,Mendosa3} were proposed: the volumetric approach which exactly conserves the mass and momentum without a mass source, and also the curvilinear body force which adds momentum in the curvilinear space leads to the exact momentum conservation and reproduces the Navier-Stokes hydrodynamics to the viscous order. 

The goal of this paper is to test and numerically validate this new  method on several exactly solvable fluid flow cases in which non-equilateral lattices can be used and results can be compared with both analytical solutions and standard LBM.  We have chosen the following four  2-dimensional test models: planar Couette and Poiseuille flows, with rectangular, but possibly non-equilateral, lattice cells, and circular Couette and  Poiseuille flows, with strong effects of cells' natural curvilinearity.
 
We consider two LBM lattices: D2Q9 and D2Q21 with different degrees of moment isotropy. The D2Q9 has the moment isotropy up to the 4th order, whereas the D2Q21 has the moment isotropy up to the 6th order. In the first two of the considered - rectangular yet non-equilateral - cases, for both the D2Q9 and D2Q21 lattices we observe a very good convergence of the new numerical method \cite{HChen} to the exact solution with resolution increase. In the second two cases characterized by the true naturally curvilinear lattice cells, we find that only the D2Q21 lattice leads to convergence to the exact solutions. Apparently, the D2Q9 4th-order moment isotropy is not sufficient for ensuring the accuracy of the new method, since the numerical method does converge to a solution that is however different from the exact one. Therefore, we conclude that the higher-order (6th or higher) moment isotropy of the lattice is required for the volumetric curvilinear LBM method \cite{HChen}. 

We have studied both equilibrium cases with the LBM relaxation time \(\tau=1\) and non-equilibrium cases with \(\frac{1}{2}<\tau<1\) and confirmed that the method \cite{HChen} works well for non-equilibrium cases as well. 

In order to achieve our goals, we have generalized the LBM periodicity, no-slip, and moving wall boundary conditions to the fully curvilinear case. Additionally, as explained in more detail later, we have proposed an adjustment procedure using the "no flow" solution algorithm to adjust the effects of discrete finite-difference approximation for the generalized basis vectors definition in \cite{HChen}. 

In the next section, we review the theoretical approach formulated in \cite{HChen}. Then in Section 3, we present numerical results for the four basic flow cases mentioned above. In Section 4 we further discuss our findings as well some possible future work directions. In Appendix A we discuss the properties of moments isotropy of the considered lattices D2Q9 and D2Q21. In Appendix B, we formulate the curvilinear boundary conditions. In Appendix C, we describe an example of a non-equidistant lattice used for this study.

\section{Formulation of LBM in Curvilinear Coordinates} 

In \cite{HChen}, a volumetric lattice Boltzmann formulation on a general curvilinear mesh is constructed based on a one-to-one mapping between physical and computational spaces ${\bf x} = {\bf x}({q})$ as follows. The coordinate values in the computational space $\{ {q} \}$ are defined exactly as in standard LBM, i.e. forming a 3D Cartesian lattice with the lattice spacing unity.
The nearest neighbor of a site in the physical space ${\bf x}({q})$ along the $i$th ($i = 1, 2, 3$)  coordinate direction in the positive or negative direction is a spatial point ${\bf x}_{\pm i} = {\bf x}({q}_{\pm i})$, where ${q}_{\pm i}$ is a unique coordinate value for the neighboring site, so that 
${q}_{\pm i} = (q^1_{\pm i}, q^2_{\pm i}, q^3_{\pm i})$
and $q^j_{\pm i} - q^j = \pm  \delta^j_i$. This defines the distance vector from ${\bf x}({q})$ to one of its neighbors ${\bf x}({q}_{\pm i})$: 
\begin{equation} 
{\bf D}_{\pm i} ({q}) \equiv {\bf x}({q}_{\pm i}) - {\bf x}({q}); \;\;\; i = 1, 2, 3
\label{distance}
\end{equation} 
which allows construction of the basis tangent vectors at ${\bf x}({q})$: 
\begin{eqnarray}
{\bf g}_i ({q}) \equiv [{\bf D}_i ({q}) - {\bf D}_{-i}({q})]/2 \Delta x; \nonumber
\label{tangent}
\end{eqnarray} 
that have a number of standard differential geometry properties that can be found in \cite{HChen}. The metric tensor and the cell volume $J$ at ${\bf x}({q})$ are thus defined as
\begin{equation} 
g_{ij}({q}) \equiv {\bf g}_i ({q}) \cdot {\bf g}_j ({q}); \;\;
J({q}) \equiv ({\bf g}_1 ({q}) \times {\bf g}_2 ({q})) \cdot {\bf g}_3 ({q})
\label{metric}
\end{equation}
and the co-tangent basis vectors ${\bf g}^i({q})$ as well as the inverse metric tensor,
\begin{eqnarray}
{\bf g}^i({q})\equiv \epsilon^{ijk}{\bf g}_j({q}) \times {\bf g}_k({q}) / J({q});\;\;\;{\bf g}_i({q}) \cdot {\bf g}^j({q}) = \delta_i^j;\;\;\; 
g^{ij}({q}) \equiv {\bf g}^i ({q}) \cdot {\bf g}^j ({q}),
\label{covectors}
\end{eqnarray}
where \(\epsilon^{ijk}\) is a standard 3-dimensional Levy-Civita symbol. 

With these definitions, we obtain the lattice Boltzmann velocity vectors on
a general curvilinear mesh defined similar to the ones on a standard Cartesian lattice,
\begin{equation}
{\bf e}_\alpha ({q}) \equiv {c}^i_\alpha {\bf g}_i({q}) 
\Delta x/\Delta t 
\label{lattice}
\end{equation}
as well as a discrete analog of the Christoffel symbol,
\begin{eqnarray}
\Theta^i_j({q} + {c}_\alpha , {q}) 
\equiv [{\bf g}_j({q} + {c}_\alpha) - {\bf g}_j({q})] \cdot {\bf g}^i({q}). 
 \;\;\; \alpha = 0, 1, \ldots , b
\label{gamma}
\end{eqnarray}

Now, the evolution of particle distribution is defined in the computational space $q$ 
similar to the standard isothermal lattice Boltzmann equation (LBE), \cite{FHP1,SChen,Benzi,CCM,Qian}
\begin{equation}
N_\alpha({q} + {c}_\alpha, t + 1) = N_\alpha({q}, t) + \Omega_\alpha({q}, t) + \delta N_\alpha({q}, t),
\label{lbe}
\end{equation}
where $N_\alpha({q}, t)$ is the number of particles belonging to the discrete direction 
${c}_\alpha$ in the cell ${q}$ at time $t$.  Here, $\Omega_\alpha({q}, t)$ is the collision term that
satisfies local mass and momentum conservation,
and the particle density distribution function $f_\alpha({q}, t)$ is related to $N_\alpha({q}, t)$ via
\begin{equation}
J({q})f_\alpha({q}, t) = N_\alpha({q}, t).
\label{dist}
\end{equation}

The fundamental fluid quantities such as density $\rho ({q}, t)$ and velocity ${\bf u} ({q}, t)$ are given by the standard hydrodynamic moments,
\begin{eqnarray}
\rho ({q}, t) = \sum_\alpha f_\alpha ({q}, t);\;\;\;
\rho ({q}, t) {\bf u} ({q}, t)= \sum_\alpha {\bf e}_\alpha ({q}) f_\alpha ({q}, t)
\label{moments}
\end{eqnarray} 

Using the Eq. \eqref{lattice}, the velocity moment above can  be rewritten as 

\begin{eqnarray}
\rho ({q}, t) {\bf u} ({q}, t) = \sum_\alpha {c}^i_\alpha {\bf g}_i({q}) f_\alpha ({q}, t)
=\rho ({q}, t) {U}^i ({q}, t) {\bf g}_i({q})
\label{velocity}
\end{eqnarray}
and the velocity in the curvilinear coordinate system is given by:
\begin{equation}
\rho ({q}, t) {U}^i ({q}, t) = \sum_\alpha {c}^i_\alpha f_\alpha ({q}, t);\;\;
\rho ({q}, t) {U} ({q}, t) = \sum_\alpha {c}_\alpha f_\alpha ({q}, t). 
\label{vcomp}
\end{equation}
Observe that the Eq. \eqref{vcomp} has the same form for the fluid velocity
as that in the standard Cartesian lattice-based LBM. We will use a linearized LBM collision term \cite{Benzi,Molvig}:
\begin{equation}
\Omega_\alpha({q}, t) = -J({q})[f_\alpha ({q}, t) - f^{eq}_\alpha ({q}, t)]/\tau;\;\;\;
\label{collide}
\end{equation}
where $f^{eq}_\alpha ({q}, t)$ is the equilibrium distribution function and $\tau$  is the Bhatnagar-Gross-Krook (BGK) collision relaxation time \cite{BGK,SChen,CCM,Qian}.  

The extra term $\delta N_\alpha({q}, t)$ in the Eq. \eqref{lbe} represents
the change of particle distribution due to an effective inertial body force, which is a key feature of curvilinear geometry-based LBM, associated with the curvature and non-uniformity of a general curvilinear mesh. This inertial body force obviously vanishes in the standard LBM on a Cartesian lattice.

Define the advection process as an exact one-to-one hop from one site in the computational space \(\vec{q}\) to another as in the standard
LBM:
\begin{equation} 
N_\alpha({q} + {c}_\alpha, t + 1) = N'_\alpha({q}, t), 
\label{adv}
\end{equation}
where $N'_\alpha({q}, t)$ is the post-collide distribution at $({q}, t)$ that is equal to the right side of the Eq. \eqref{lbe}. In \cite{HChen} it was shown that the key intrinsic effect of curvilinear formulation, the net momentum change via advection 
from all the neighboring cells into cell ${q}$ is given by: 
\begin{equation}
J({q}) {\bm{\chi}}^I({q},t) = 
- \sum_{\alpha} [{\bf e}_{\alpha} ({q}) - {\bf e}_{\alpha} ({q} - {c}_{\alpha})]N_{\alpha}({q},t), 
\label{forceI}
\end{equation}
and out of cell ${q}$ to all its neighboring cells, is given by: 
\begin{equation}
J({q}) {\bm{\chi}}^o({q},t) = 
- \sum_{\alpha} [{\bf e}_{\alpha} ({q} + {c}_{\alpha}) - {\bf e}_{\alpha} ({q})]N'_{\alpha}({q},t)
\label{forceO}
\end{equation}
so that an ``inertial force'' \(\bm{\chi}({q},t)=\left[\bm{\chi}^I({q},t)+\bm{\chi}^o({q},t)\right]/2\) that equals exactly to the amount needed for achieving the momentum conservation in the underlying Euclidean space is:
\begin{eqnarray}
{F}^i({q},t)={\bm{\chi}}({q},t)\cdot{\bf g}^i({q}) \nonumber \\
=-\frac{1}{2J({q})}
\sum_{\alpha}c^j_{\alpha} \{\Theta^i_j({q} + {c}_{\alpha},{q}) N'_{\alpha}({q},t)-\Theta^i_j({q}-{c}_{\alpha}, {q})N_{\alpha}({q},t)\}. 
\label{compo-constr}
\end{eqnarray}

The full viscous Navier-Stokes equation is recovered when the momentum flux is defined as: 
\begin{equation}
\delta{\Pi}^{ij}({q}, t) \equiv - \frac{1}{2}\left(1-\frac{1}{2\tau}\right) \sum_\alpha {c}^i_\alpha {c}^k_\alpha [\Theta^j_k({q} + {c}_\alpha , {q}) - \Theta^j_k({q} - {c}_\alpha , {q})] f^{eq}_\alpha({q}, t). 
\label{delta-flux}
\end{equation}
It is shown in \cite{HChen} that choosing 
\begin{equation}
\delta N_\alpha({q}, t) = w_{\alpha}J({q}) [\frac {{c}^j_\alpha {F}^j({q}, t)} {T_0}
+ (\frac {{c}^j_\alpha {c}^k_\alpha} {T_0} - \delta^{jk}) \frac {\delta{\Pi}^{jk}({q}, t)} {T_0} ]
\label{deltaN}
\end{equation}
satisfies the necessary moment constraints. Note also that due to the appearance of $N'_\alpha({q}, t)$ in the Eq. \eqref{compo-constr}, the overall collision process for determining $N'_\alpha({q}, t)$ defines an implicit relationship. Specifically, the Eq. \eqref{deltaN} defines the curvilinear correction \(\delta N_{\alpha}\left(q,t\right)\) through \(F^j \left(q,t\right) \) and the Eq. \eqref{compo-constr} expresses \(F^j \left(q,t\right) \) using \(N'\left(q,t\right)\), which, again, depends on the same \(\delta N_{\alpha}\left(q,t\right)\). In this work, for the numerical implementation of the Eq. \eqref{compo-constr} we split this implicit relationship into explicit relationships at two successive time steps: we use the \(N'_{\alpha}\left(q,t-1\right)\) from the previous time-step in the Eq. \eqref{compo-constr}. Iterative procedures to numerically implement this implicit relationship in the Eq. \eqref{compo-constr} are also possible. 

The physical velocity \(\tilde{U}\) was defined in \cite{HChen} through the curvilinear body force \(F\) as follows: 
\begin{equation}
\tilde{U}^i(q,t)=U^i(q,t)+\frac{a^i(q,t)}{2}, \mbox{ where: } a^i(q,t)\equiv\frac{F^i(q,t)}{\rho(q,t)}. 
\end{equation}

The equilibrium distribution function that produces the Navier-Stokes equation in curvilinear coordinates in the hydrodynamic limit is \cite{HChen}: 
\begin{eqnarray}
f^{eq}_\alpha = & & \rho w_\alpha \bigg\{ 1 + \frac {{c}^i_\alpha {U}^i} {T_0}
+ \frac {1} {2T_0} (\frac {{c}^i_\alpha {c}^j_\alpha} {T_0} - \delta^{ij})
[(g^{ij} - \delta^{ij})T_0 + {\tilde U}^i {\tilde U}^j] \nonumber \\
& & + \frac {1} {6T_0^3} ({c}^i_\alpha {c}^j_\alpha {c}^k_\alpha 
- T_0 ({c}^i_\alpha \delta^{jk} + {c}^j_\alpha \delta^{ki} + {c}^k_\alpha \delta^{ij}))
[ T_0 [ (g^{ij}{\tilde U}^k - \delta^{ij}U^k) \nonumber \\
& & + (g^{jk}{\tilde U}^i - \delta^{jk}U^i) + (g^{ki}{\tilde U}^j - \delta^{ki}U^j)]
+ {\tilde U}^i {\tilde U}^j {\tilde U}^k ]  \bigg\}. 
\label{equilibrium}
\end{eqnarray}

We will also need the following simple mapping of fluid values $\rho ({q}, t)$ and ${U}^i({q}, t)$ onto the original curvilinear mesh:
\begin{equation}
    \rho ({\bf x}({q}), t) = \rho ({q}, t); \;\;\;  
    {\bf u}({\bf x}({q}), t) = {U}^i({q}, t) {\bf g}_i ({q}). 
\label{result}
\end{equation}

\section{Four Exactly Solvable Flow Problems}
The four exactly solvable flow problems are presented below in order to validate the new approach \cite{HChen}. As stated in \cite{HChen}, in order to recover the correct isothermal low Ma Navier-Stokes hydrodynamics, a set of necessary moments isotropy and normalization conditions must be satisfied. This is why we present a detailed comparison between two lattices: D2Q9 and D2Q21. The details of isotropy and normalization conditions for these two lattices are presented in Appendix A. In all of the below cases, we use the boundary conditions, generalized by us for curvilinear LBM, presented in detail in Appendix B. 

\subsection{Planar Couette Flow}

This well-known \cite{Batchelor} simple exact solution \( \vec{u}(\vec{x},t)= (u_x, u_y) \) describes a stationary flow of viscous incompressible fluid between two vertical planes: at \( x=0 \) moving with a constant velocity \( \overrightarrow{U}_0=  (0, -U_0)\), and a non-moving one at \( x=l \): 
\begin{equation}
\label{PlaneCouetteExactSolution}
u_y=U_0\left(\frac{x}{l}-1\right). 
\end{equation}

Solution of the Eq. \eqref{PlaneCouetteExactSolution} in lattice units for a non-equidistant lattice is: 
\begin{equation}
\label{PlaneCouetteSolutionLatticeUnits}
u_{y,i}^{lat}=U_{0}^{lat}\left(\frac{x_i}{\overline{\triangle}N_x}-1\right).
\end{equation}
Here \(x_i\) is a spatially varying coordinate of the lattice node. In this study, a lattice that is linearly contracting towards the boundaries is used (see Appendix C), and \(\overline{\triangle}\) is the average step \(\overline{\triangle}=l/N_x\) in the \(x\)-direction. We studied other variable-size lattices and conclude that our main findings do not depend on the  particular type of contraction, as long as some general stability conditions are satisfied.

Thus defined plane channel problem does not possess a true curvilinearity but rather a deviation from the equidistant lattice which is due to cell contraction. 

The LBM parameters for the equilibrium case with \(\tau=1\) for the D2Q9 lattice are listed in \Cref{PCTau1}, \Cref{PCD2Q9}: 
\begin{table}[H]
\centering
\caption{Planar Couette LBM parameters}
\label{PCTau1}
\begin{subtable}{\hsize}
\centering
\subcaption{D2Q9 lattice, variable \(N_x\), \(\tau=1\)}
\label{PCD2Q9}
\scalebox{0.93}{
\begin{tabular}{|p{0.01cm}|p{0.3cm}p{0.3cm}p{0.5cm}p{0.2cm}|}
\hline
\(N_x\) & \(U_{lat}\) & \(\nu_{lat}\) & \(Ma_{sim}\) & \(Re\) \\
\hline
\(8\) & \(0.208\) & \(0.167\) & \(0.361\) & \(10.0\) \\
\(16\) & \(0.208\) & \(0.167\) & \(0.361\) & \(20.0\) \\
\(32\) & \(0.208\) & \(0.167\) & \(0.361\) & \(40.0\) \\
\(64\) & \(0.208\) & \(0.167\) & \(0.361\) & \(80.0\) \\
\(128\) & \(0.208\) & \(0.167\) & \(0.361\) & \(160.0\) \\
\hline
\end{tabular}}
\end{subtable}
\begin{subtable}{\hsize}
\centering
\subcaption{D2Q21 Lattice, variable \(N_x\), \(\tau=1\)}
\label{PCD2Q21}
\scalebox{0.93}{
\begin{tabular}{|p{0.01cm}|p{0.3cm}p{0.3cm}p{0.5cm}p{0.1cm}|}
\hline
\(N_x\) & \(U_{lat}\) & \(\nu_{lat}\) & \(Ma_{sim}\) & \(Re\) \\
\hline
\(8\) & \(0.208\) & \(0.333\) & \(0.255\) & \(5.0\) \\
\(16\) & \(0.208\) & \(0.333\) & \(0.255\) & \(10.0\) \\
\(32\) & \(0.208\) & \(0.333\) & \(0.255\) & \(20.0\) \\
\(64\) & \(0.208\) & \(0.333\) & \(0.255\) & \(40.0\) \\
\(128\) & \(0.208\) & \(0.333\) & \(0.255\) & \(80.0\) \\
\hline
\end{tabular}}
\end{subtable}
\begin{subtable}{\hsize}
\centering
\subcaption{D2Q9 Lattice, variable \(\tau\), \(N_x=64\)}
\label{PCD2Q9N64}
\scalebox{0.93}{
\begin{tabular}{|p{0.05cm}|p{0.3cm}p{0.3cm}p{0.5cm}p{0.1cm}|}
\hline
\(\tau\) & \(U_{lat}\) & \(\nu_{lat}\) & \(Ma_{sim}\) & \(Re\) \\
\hline
\(1.0\) & \(0.052\) & \(0.167\) & \(0.090\) & \(20.0\) \\
\(0.9\) & \(0.042\) & \(0.133\) & \(0.072\) & \(20.0\) \\
\(0.8\) & \(0.031\) & \(0.100\) & \(0.054\) & \(20.0\) \\
\(0.7\) & \(0.021\) & \(0.067\) & \(0.036\) & \(20.0\) \\
\hline
\end{tabular}}
\end{subtable}
\begin{subtable}{\hsize}
\centering
\subcaption{D2Q21 lattice, variable \(\tau\), \(N_x=64\)}
\label{PCD2Q21N64}
\scalebox{0.93}{
\begin{tabular}{|p{0.05cm}|p{0.3cm}p{0.3cm}p{0.5cm}p{0.1cm}|}
\hline
\multicolumn{5}{|c|}{Planar Couette, D2Q21 Lattice, \(N_x=64\)} \\
\hline
\(\tau\) & \(U_{lat}\) & \(\nu_{lat}\) & \(Ma_{sim}\) & \(Re\) \\
\hline
\(1.0\) & \(0.052\) & \(0.333\) & \(0.064\) & \(10.0\) \\
\(0.9\) & \(0.042\) & \(0.267\) & \(0.051\) & \(10.0\) \\
\(0.8\) & \(0.031\) & \(0.200\) & \(0.038\) & \(10.0\) \\
\(0.7\) & \(0.021\) & \(0.133\) & \(0.026\) & \(10.0\) \\
\hline
\end{tabular}}
\end{subtable}
\end{table}
As it is well expected, on the equidistant D2Q9 lattice (which corresponds to the trivial compression ratio \(CR=0\) in our notations), all quantities \(\rho(x)\), \(\tilde{U}^1(x)\) and \(\tilde{U}^2(x)\) for all resolutions  \(N_x=8,16,32,64,128\) accurately reproduce the exact analytical solution Eqs.  \mref{PlaneCouetteSolutionLatticeUnits, PlaneCouetteExactSolution}. 

\begin{figure}[H]
\centering
\caption{Examples of Geometries and Lattices Considered}
\begin{subfigure}[H]{0.6\textwidth}
\subcaption{Cartesian geometry case with \(N_x=N_y=32\), and linear \(CR=0.4\) grid compression in \(x-\)direction}
\label{GridContractionPlanarGeometry}
\includegraphics[width=\textwidth]{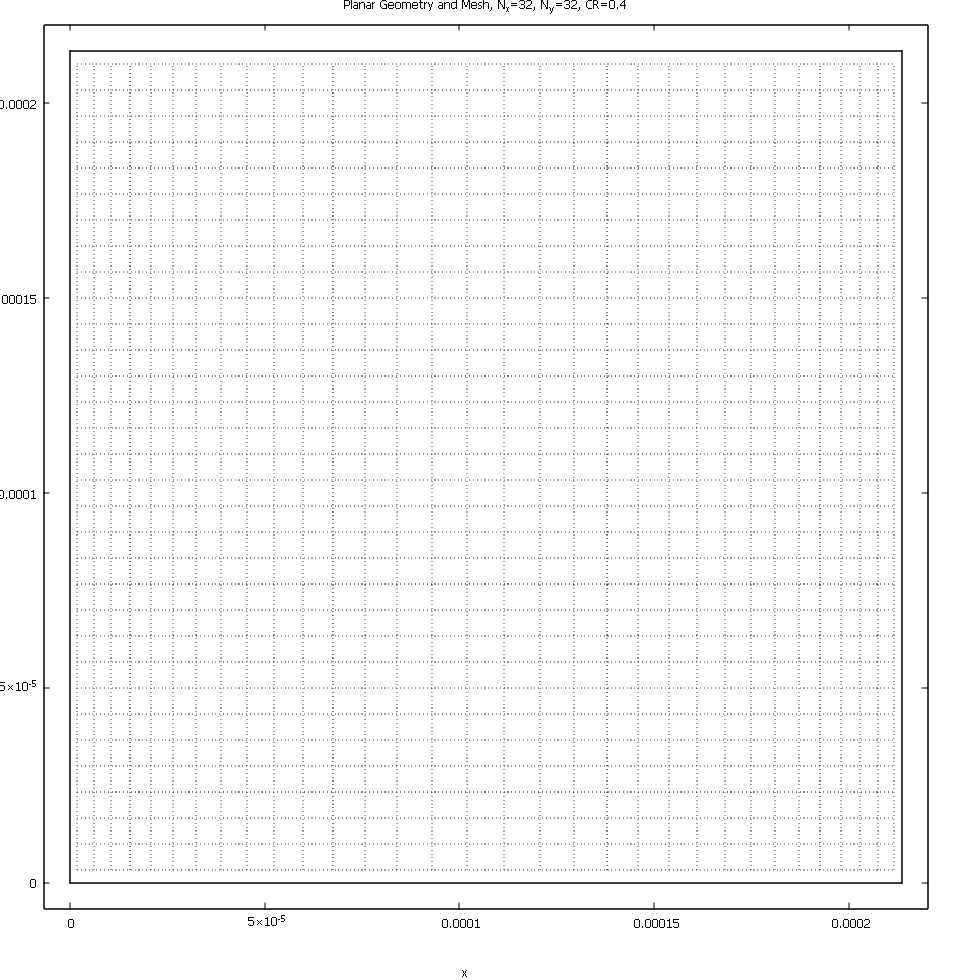}
\end{subfigure}
\begin{subfigure}[H]{0.6\textwidth}
\centering
\subcaption{Circular geometry case with \(N_r=32, N_{\theta}=20\) and \(NCR=11\)}
\label{CircularGeometryNCR11}
\includegraphics[width=\textwidth]{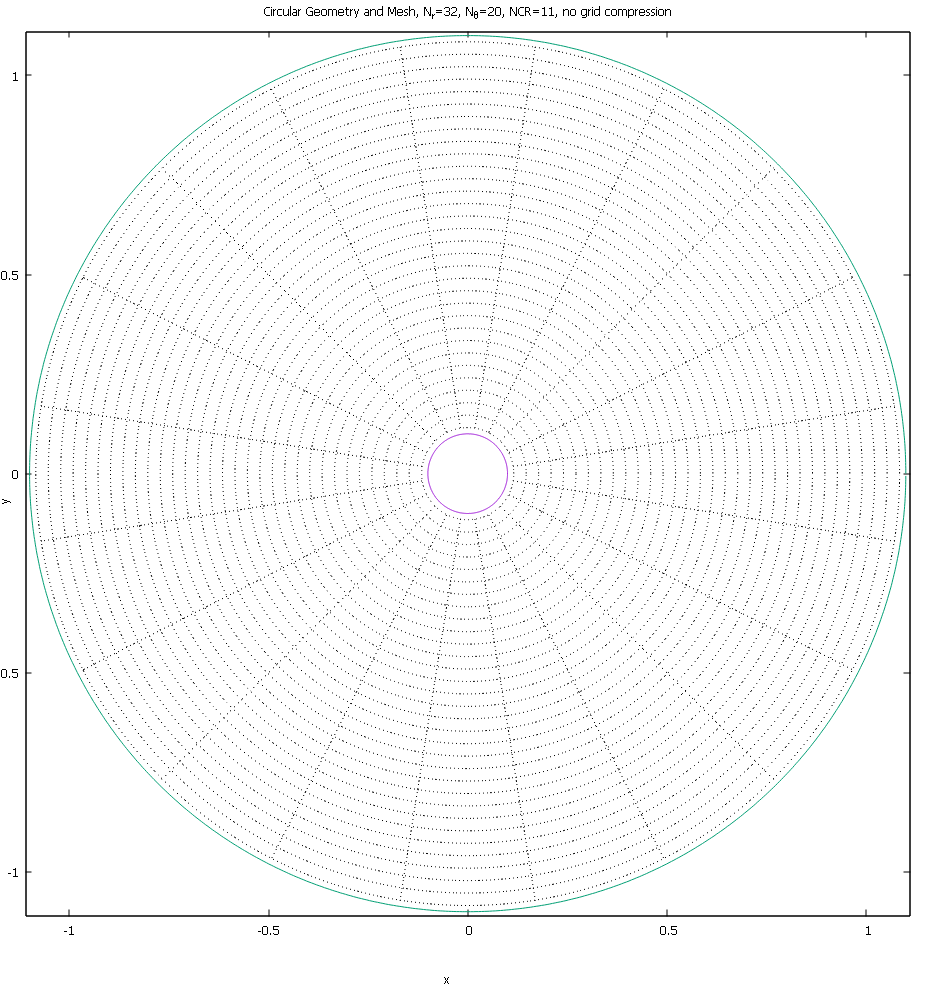}
\end{subfigure}
\end{figure}

One simple way to introduce some curvilinear effects into an otherwise Cartesian geometry is to consider variable aspect ratio grid cells, for example as is done in the linear grid compression case described in Appendix C. An example of such geometry and grid is shown in \Cref{GridContractionPlanarGeometry}. 
The numerical results for the fields of \(\rho(x)\) and \(\tilde{U}^2(x)\) for the D2Q9 lattice with the nontrivial compression ratio \(CR=0.4\), are presented in Figs. \mref{FigPCD2Q9Rho,FigPCD2Q9Ut2} below:
\begin{figure}[H]
\centering
\caption{Planar Couette flow, contracting grid, variable \(N_x\), D2Q9 lattice}
\begin{subfigure}[H]{0.6\textwidth}
\centering
\subcaption{\(\rho(x)\) for \(N_x=8,16,32,64,128\)  and \(CR=0.4\)}
\label{FigPCD2Q9Rho}
\includegraphics[width=\textwidth]{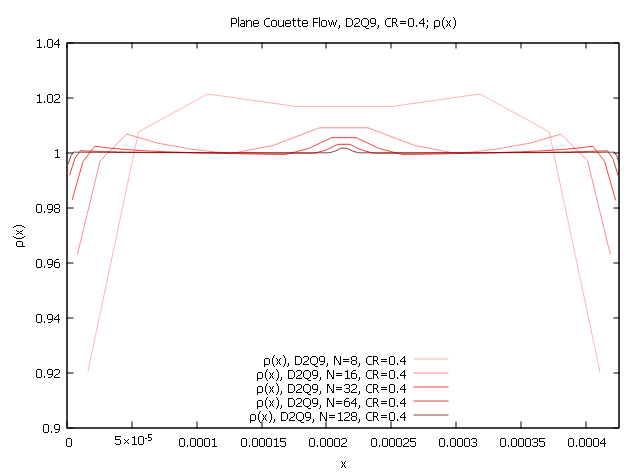}
\end{subfigure}
\begin{subfigure}[H]{0.6\textwidth}
\centering
\subcaption{\(\tilde{U}^2(x)\) for \(N_x=8,16,32,64,128\) and \(CR=0.4\)}
\label{FigPCD2Q9Ut2}
\includegraphics[width=\textwidth]{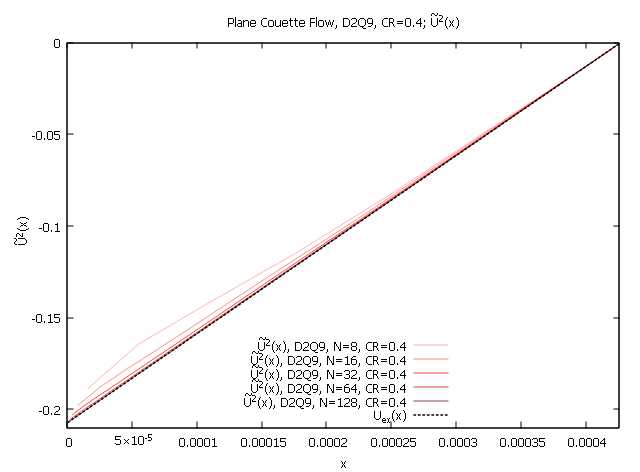}
\end{subfigure}
\end{figure}

\(\tilde{U}^1(x)\) in this case is negligibly small for all resolutions \(N_x=8,16,32,64,128\). 
As we can see, even on strongly non-equidistant lattice D2Q9 we also converge to the exact solution with higher resolutions.

The LBM parameters for the D2Q21 lattice and \(\tau=1\) that we used are listed in the \Cref{PCTau1}, \Cref{PCD2Q21} above. 

The boundary conditions developed for curvilinear LBM with D2Q21 are presented in Appendix B. In formulating the boundary conditions, we have assumed the symmetrically-continued geometry through the boundary. The stencil length for D2Q21 is three times larger than that for D2Q9, and therefore some approximation errors are expected to be larger near the moving boundary than those for D2Q9. It needs to be pointed out that this issue is related to the boundary conditions algorithm rather than to the intrinsic nature of the curvilinear LBM \cite{HChen}. 

Similar to the D2Q9 case, our code applied to the trivial equilateral D2Q21 lattice (\(CR=0\)) gives perfectly converging results for all \(\rho(x)\), \(\tilde{U}^1(x)\), and \(\tilde{U}^2(x)\), for all resolutions  \(N_x=8,16,32,64,128\) that accurately reproduce the exact analytical solution Eqs. \mref{PlaneCouetteSolutionLatticeUnits,PlaneCouetteExactSolution}. Results for \(\rho(x)\) and \(\tilde{U}^2(x)\) for D2Q21 and nontrivial rectangular lattice with \(CR=0.4\) are presented in Figs. \ref{FigPCD2Q21Rho}, \ref{FigPCD2Q21Ut2} below. Note that \(CR=0.4\) corresponds to \(\triangle_1=0.6\times\overline{\triangle} \) and \(\triangle_{\frac{N}{2}}=2.33\times\triangle_1\), and that according to our set-up of the linearly contracting lattice described in Appendix C, this value of \(CR\) corresponds to different values of lattice parameter \(a\) for each resolution. 

\begin{figure}[H]
\centering
\caption{Planar Couette flow, contracting grid, variable \(N_x\), D2Q21 lattice}
\begin{subfigure}[H]{0.6\textwidth}
\centering
\subcaption{\(\rho(x)\) for \(N_x=8,16,32,64,128\)  and \(CR=0.4\)}
\label{FigPCD2Q21Rho}
\includegraphics[width=\textwidth]{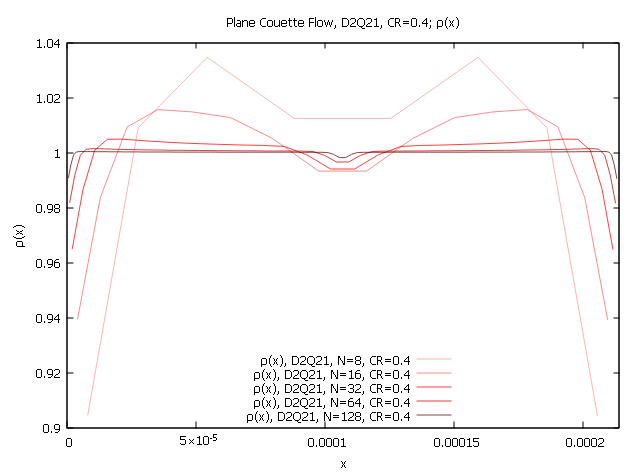}
\bigskip
\end{subfigure}
\begin{subfigure}[H]{0.6\textwidth}
\centering
\subcaption{\(\tilde{U}^2(x)\) for \(N_x=8,16,32,64,128\) and \(CR=0.4\)}
\label{FigPCD2Q21Ut2}
\includegraphics[width=\textwidth]{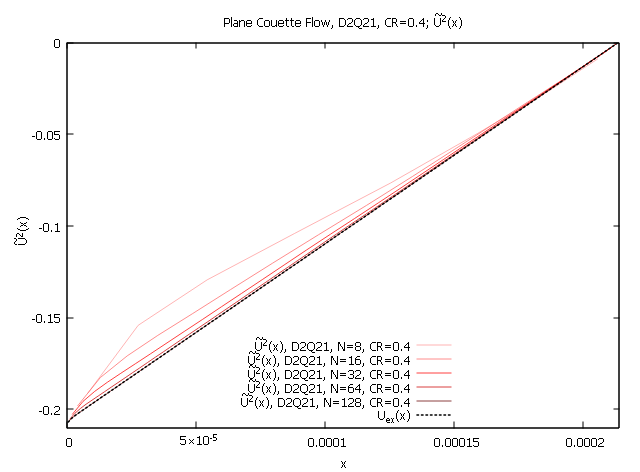}
\end{subfigure}
\end{figure}
As we can see, even on strongly non-equidistant lattice D2Q21 we also converge to the exact solution with higher resolutions. 

Let us now present the results of numerical model solutions for a fixed resolution \(N_x=64\) but variable \(\tau=\{1.0,0.9,0.8,0.7\}\) for both lattices D2Q9 and D2Q21. 

The corresponding LBM parameters for the D2Q9 lattice we used are listed in \Cref{PCTau1}, \Cref{PCD2Q9N64} above. 

\begin{figure}[H]
\centering
\caption{Planar Couette flow, contracting grid, variable \(\tau\) for D2Q9 lattice}
\begin{subfigure}[H]{0.6\textwidth}
\centering
\subcaption{\(\frac{\overline{U}^2(x)}{U_{lat}}, \frac{U_{ex}(x)}{U_{lat}}\) for D2Q9, \(\tau=1.0,0.9,0.8,0.7\),  and \(CR=0.35\)}
\label{PCD2Q9Ut2Ulat}
\includegraphics[width=\textwidth]{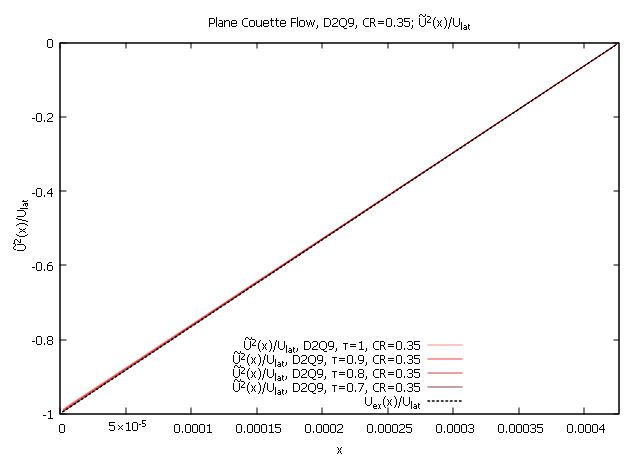}
\end{subfigure}
\begin{subfigure}[H]{0.6\textwidth}
\centering
\subcaption{\(\frac{\overline{U}^2(x)}{\overline{U}^2_{ex}(x)}\) for D2Q9, \(\tau=1.0,0.9,0.8,0.7\),  and \(CR=0.35\)}
\label{PCD2Q9Ut2Uex}
\includegraphics[width=\textwidth]{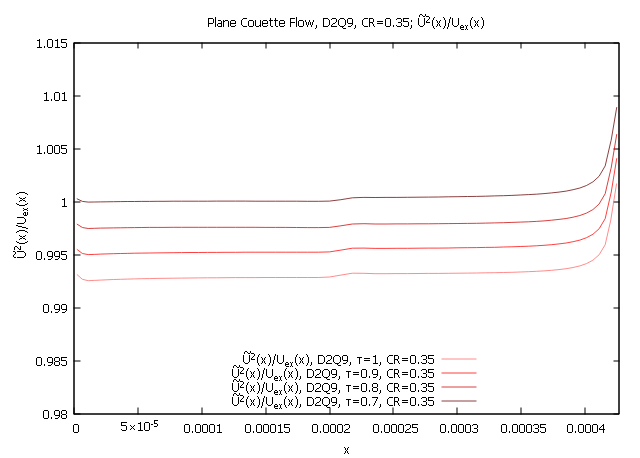}
\end{subfigure}
\end{figure}

The agreement of numerical solution with the exact one with no lattice compression \(CR=0\) is of course very good for both lattices D2Q9 and D2Q21 for all \(\tau=\{1.0,0.9,0.8,0.7\}\). 
In Figs. \ref{PCD2Q9Ut2Ulat}, \ref{PCD2Q9Ut2Uex} below we present the comparisons of numerical results with the exact solution for lattice D2Q9 with compression \(CR=0.35\) and \(\tau=\{1.0,0.9,0.8,0.7\}\). We show variable \(\tau\) cases on the same graphs, for the following quantities: \(\frac{\overline{U}^2(x)}{U_{lat}}, \frac{U_{ex}(x)}{U_{lat}}\), and \(\frac{\overline{U}^2(x)}{\overline{U}^2_{ex}(x)}\). 
The \(\frac{\overline{U}^1(x)}{U_{lat}}\) for D2Q9, \(\tau=1.0,0.9,0.8,0.7\),  and \(CR=0.35\), are all equal to zero within double-precision accuracy and the \(\frac{\rho(x)}{\rho_0}\) for D2Q9, \(\tau=1.0,0.9,0.8,0.7\),  and \(CR=0.35\) show very weak dependence on \(\tau\). We can see that for D2Q9 lattice and strong cell compression the new method \cite{HChen} not only converges with the increase in resolution but also works well for various values of parameter \(\tau\). 

The corresponding LBM parameters used for the D2Q21 lattice are listed in \Cref{PCTau1}, \cref{PCD2Q21N64} above. Similar numerical results for D2Q21 lattice are presented 
 in \mref{PCD2Q21Ut2Ulat,PCD2Q21Ut2Uex} below. 
\begin{figure}[H]
\centering
\caption{Planar Couette flow, contracting grid, variable \(\tau\) for D2Q21 lattice}
\begin{subfigure}[H]{0.6\textwidth}
\centering
\subcaption{\(\frac{\overline{U}^2(x)}{U_{lat}}, \frac{U_{ex}(x)}{U_{lat}}\) for D2Q21, \(\tau=1.0,0.9,0.8,0.7\),  and \(CR=0.35\)}
\label{PCD2Q21Ut2Ulat}
\includegraphics[width=\textwidth]{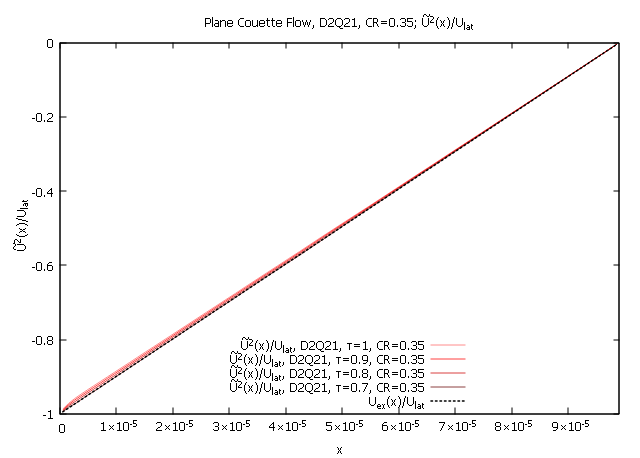}
\end{subfigure}
\begin{subfigure}[H]{0.6\textwidth}
\centering
\caption{\(\frac{\overline{U}^2(x)}{\overline{U}^2_{ex}(x)}\) for D2Q21, \(\tau=1.0,0.9,0.8,0.7\),  and \(CR=0.35\)}
\label{PCD2Q21Ut2Uex}
\includegraphics[width=\textwidth]{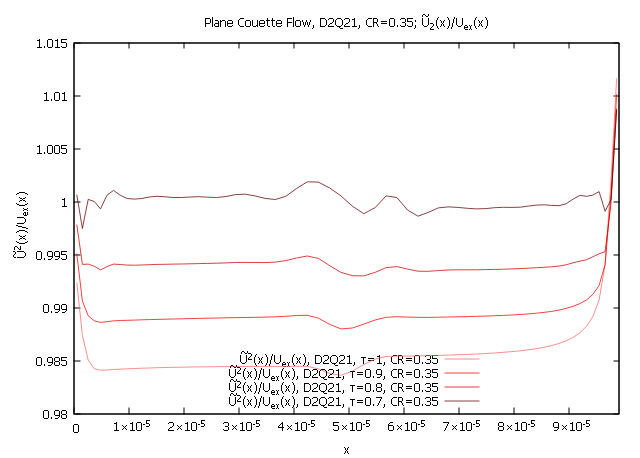}
\end{subfigure}
\end{figure}
As in the D2Q9 case, the \(\frac{\overline{U}^1(x)}{U_{lat}}\) for D2Q21, \(\tau=1.0,0.9,0.8,0.7\),  and \(CR=0.35\), are approximately equal to zero and the \(\frac{\rho(x)}{\rho_0}\) for D2Q21, \(\tau=1.0,0.9,0.8,0.7\),  and \(CR=0.35\) show very weak dependence on \(\tau\). Here we can also see that for D2Q21 lattice  with strong cell compression  the new method \cite{HChen} not only converges with the increase in resolution but also works well for various values of parameter \(\tau\). 

All the problems considered in the paper were solved numerically as non-stationary problems converging to a steady state from zero initial velocity state. The convergence to the steady state was judged by the conservation of total kinetic energy in the system in the first two plane geometry problems and by the conservation of both total kinetic energy and total angular momentum in the system in the second two curvilinear (circular) problems. In order to achieve that steady state in some cases over 150 times to traverse the characteristic length with the characteristic velocity were required. Total mass in the system was conserved at all intermediary times. 

Also note the fact that in curvilinear and non-equidistant step cases, the numerical solution for density \(\rho(\vec{q})\) is not constant, as can be seen in Figs. \ref{FigPCD2Q9Rho},\ref{FigPCD2Q21Rho}, is a byproduct of the finite-difference approximation of the basis vectors in \cite{HChen} and can be eliminated and reduced to the exact solution (\( \rho(\vec{q})=\rho_0\) in this case) using the following "no flow" adjustment: 
\begin{equation}
J\left(\vec{q}\right)\rightarrow J\left(\vec{q}\right)\frac{\rho_{nf}(\vec{q})}{\rho_0},
\end{equation}
where \(\rho_{nf}\left(\vec{q}\right)\) is the numerical solution of the corresponding "no flow" problem with the same geometry but \(U_{lat}=0\). 

The differences between D2Q9 and D2Q21 lattices do not seem to affect the convergence to the exact solution for this simple model problem. 

\subsection{Plane Poiseuille Flow}

This next well-known \cite{Batchelor} classical solution describes a stationary flow of viscous liquid between two infinite vertical planes at \(x=0\) and \(x=l\) under the action of constant vertical gravity force corresponding to acceleration \(-g\).  The stationary solution with no-slip boundary conditions at \(x=0,l\) is a parabola,  
\begin{equation}
\label{PlanePoiseuilleSolution2}
u_y=\frac{G}{2}x(x-l),
\end{equation}
where $\nu$ is the kinematic viscosity and $G\equiv g/\nu$. 
Rewritten in lattice units on a non-equidistant lattice, the Eq. \eqref{PlanePoiseuilleSolution2} becomes: 
\begin{equation}
\label{PlanePoiseuilleSolutionLatticeUnits}
u_{y,i}^{lat}=\frac{G^{lat}}{2}\frac{x_i}{\overline{\triangle}}\left(\frac{x_i}{\overline{\triangle}}-N_x\right). 
\end{equation}

Similar to the Planar Couette flow, we have investigated both lattices D2Q9 and D2Q21 and a set of resolutions \(N_x=8, 16, 32, 64, 128\) with and without lattice  contraction. 

In \Cref{PPTau1}, \Cref{PPD2Q9} below we list the set of LBM parameters used for the D2Q9 lattice: 
\begin{table}[H]
\centering
\caption{Planar Poiseuille LBM parameters}
\label{PPTau1}
\begin{subtable}{\hsize}
\centering
\subcaption{D2Q9 lattice, variable \(N_x\), \(\tau=1\)}
\label{PPD2Q9}
\scalebox{0.93}{
\begin{tabular}{|p{0.01cm}|p{0.01cm}p{0.01cm}p{1.7cm}p{0.01cm}p{0.01cm}p{0.3cm}|}
\hline
\(N_x\) & \(U_{lat}\) & \(\nu_{lat}\) & \(g_{lat}\) & \(Ma_{sim}\) & \(Re\) & \(Fr\) \\
\hline
\(8\) & \(0.208\) & \(0.167\) & \(6.51\times 10^{-3}\) & \(0.361\) & \(10.0\) & \(0.9\) \\
\(16\) & \(0.208\) & \(0.167\) & \(1.63\times 10^{-3}\) & \(0.361\) & \(20.0\) & \(1.3\) \\
\(32\) & \(0.208\) & \(0.167\) & \(4.07\times 10^{-4}\) & \(0.361\) & \(40.0\) & \(1.8\) \\
\(64\) & \(0.208\) & \(0.167\) & \(1.02\times 10^{-4}\) & \(0.361\) & \(80.0\) & \(2.6\) \\
\(128\) & \(0.208\) & \(0.167\) & \(2.54\times 10^{-5}\) & \(0.361\) & \(160.0\) & \(3.7\) \\
\hline
\end{tabular}}
\end{subtable}
\begin{subtable}{\hsize}
\centering
\subcaption{D2Q21 lattice, variable \(N_x\), \(\tau=1\)}
\label{PPD2Q21}
\scalebox{0.93}{
\begin{tabular}{|p{0.01cm}|p{0.01cm}p{0.01cm}p{1.7cm}p{0.01cm}p{0.01cm}p{0.3cm}|}
\hline
\(N_x\) & \(U_{lat}\) & \(\nu_{lat}\) & \(g_{lat}\) & \(Ma_{sim}\) & \(Re\) & \(Fr\) \\
\hline
\(8\) & \(0.208\) & \(0.333\) & \(1.30\times 10^{-2}\) & \(0.255\) & \(5.0\) & \(0.6\) \\
\(16\) & \(0.208\) & \(0.333\) & \(3.26\times 10^{-3}\) & \(0.255\) & \(10.0\) & \(0.9\) \\
\(32\) & \(0.208\) & \(0.333\) & \(8.14\times 10^{-4}\) & \(0.255\) & \(20.0\) & \(1.3\) \\
\(64\) & \(0.208\) & \(0.333\) & \(2.03\times 10^{-4}\) & \(0.255\) & \(40.0\) & \(1.8\) \\
\(128\) & \(0.208\) & \(0.333\) & \(5.09\times 10^{-5}\) & \(0.255\) & \(80.0\) & \(2.6\) \\
\hline
\end{tabular}}
\end{subtable}
\begin{subtable}{\hsize}
\centering
\subcaption{D2Q9 lattice, variable \(\tau\), \(N=64\)}
\label{PPD2Q9N64}
\scalebox{0.93}{
\begin{tabular}{|p{0.01cm}|p{0.01cm}p{0.01cm}p{1.7cm}p{0.01cm}p{0.01cm}p{0.3cm}|}
\hline
\(\tau\) & \(U_{lat}\) & \(\nu_{lat}\) & \(g_{lat}\) & \(Ma_{sim}\) & \(Re\) & \(Fr\) \\
\hline
\(1.0\) & \(0.052\) & \(0.167\) & \(2.54\times 10^{-5}\) & \(0.090\) & \(20.0\) & \(1.3\) \\
\(0.9\) & \(0.042\) & \(0.133\) & \(1.63\times 10^{-5}\) & \(0.072\) & \(20.0\) & \(1.3\) \\
\(0.8\) & \(0.031\) & \(0.100\) & \(9.16\times 10^{-6}\) & \(0.054\) & \(20.0\) & \(1.3\) \\
\(0.7\) & \(0.021\) & \(0.067\) & \(4.07\times 10^{-6}\) & \(0.036\) & \(20.0\) & \(1.3\) \\
\hline
\end{tabular}}
\end{subtable}
\begin{subtable}{\hsize}
\centering
\subcaption{D2Q21 lattice, variable \(\tau\), \(N=64\)}
\label{PPD2Q21N64}
\scalebox{0.93}{
\begin{tabular}{|p{0.01cm}|p{0.01cm}p{0.01cm}p{1.7cm}p{0.01cm}p{0.01cm}p{0.3cm}|}
\hline
\(\tau\) & \(U_{lat}\) & \(\nu_{lat}\) & \(g_{lat}\) & \(Ma_{sim}\) & \(Re\) & \(Fr\) \\
\hline
\(1.0\) & \(0.052\) & \(0.333\) & \(5.09\times 10^{-5}\) & \(0.064\) & \(10.0\) & \(0.9\) \\
\(0.9\) & \(0.042\) & \(0.267\) & \(3.26\times 10^{-5}\) & \(0.051\) & \(10.0\) & \(0.9\) \\
\(0.8\) & \(0.031\) & \(0.200\) & \(1.83\times 10^{-5}\) & \(0.038\) & \(10.0\) & \(0.9\) \\
\(0.7\) & \(0.021\) & \(0.133\) & \(8.14\times 10^{-6}\) & \(0.026\) & \(10.0\) & \(0.9\) \\
\hline
\end{tabular}}
\end{subtable}
\end{table}

For D2Q9 lattice with \(CR=0\), all quantities \(\rho(x), \tilde{U}^1(x),\) and \(\tilde{U}^2(x)\) produced by the new method \cite{HChen} for all resolutions \(N_x=8,16,32,64,128\) accurately reproduce the exact analytical solution Eqs. \mref{PlanePoiseuilleSolution2,PlanePoiseuilleSolutionLatticeUnits}. 

The \Cref{PPTau1}, \Cref{PPD2Q21} above details the LBM parameters we used for the D2Q21 lattice cases. For D2Q21 lattice with \(CR=0\) the new curvilinear algorithm \cite{HChen} also converges well to the exact solution for all considered resolutions \(N_x=8,16,32,64,128\). As in the previous Couette flow case for D2Q21 we observe some very small influence of the boundary conditions. 

\begin{figure}[H]
\centering
\caption{Plane Poiseuille flow, contracting grid, variable \(N_x\), D2Q9 lattice}
\begin{subfigure}[H]{0.6\textwidth}
\centering
\subcaption{\(\rho(x)\) for \(N_x=8,16,32,64,128\)  and \(CR=0.4\)}
\label{PPD2Q9Rho}
\includegraphics[width=\textwidth]{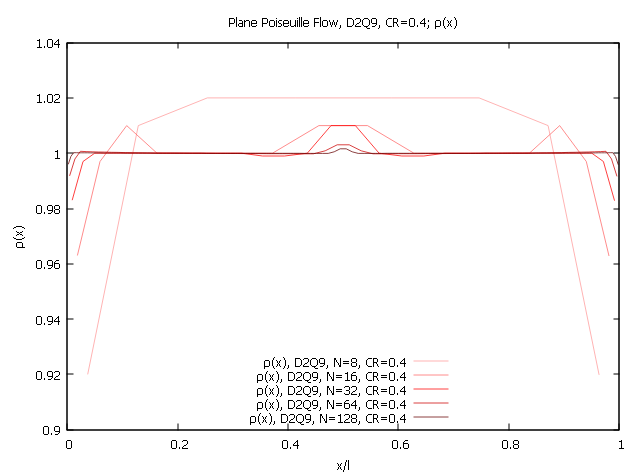}
\end{subfigure}
\begin{subfigure}[H]{0.6\textwidth}
\centering
\subcaption{\(\tilde{U}^2(x)\) for \(N_x=8,16,32,64,128\)  and \(CR=0.4\)}
\label{PPD2Q9Ut2}
\includegraphics[width=\textwidth]{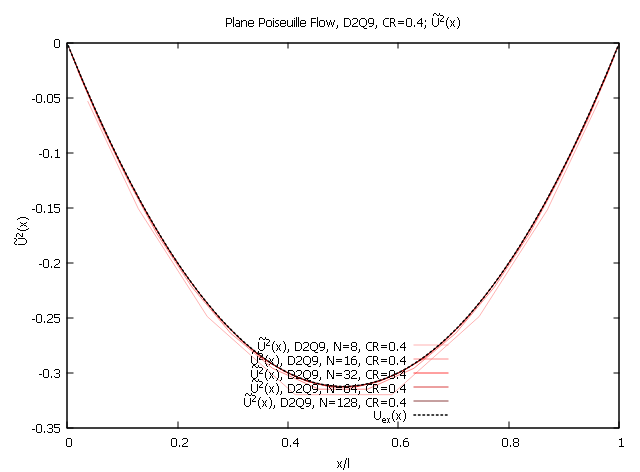}
\end{subfigure}
\end{figure}

\begin{figure}[H]
\centering
\caption{Plane Poiseuille flow, contracting grid, variable \(N_x\), D2Q21 lattice}
\begin{subfigure}[H]{0.6\textwidth}
\centering
\subcaption{\(\rho(x)\) for \(N_x=8,16,32,64,128\)  and \(CR=0.4\)}
\label{PPD2Q21Rho}
\includegraphics[width=\textwidth]{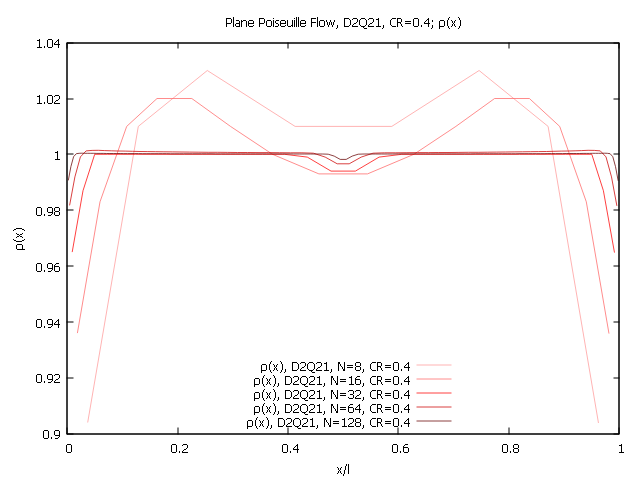}
\end{subfigure}
\begin{subfigure}[H]{0.6\textwidth}
\centering
\caption{\(\tilde{U}^2(x)\) for \(N_x=8,16,32,64,128\) and \(CR=0.4\)}
\label{PPD2Q21Ut2}
\includegraphics[width=\textwidth]{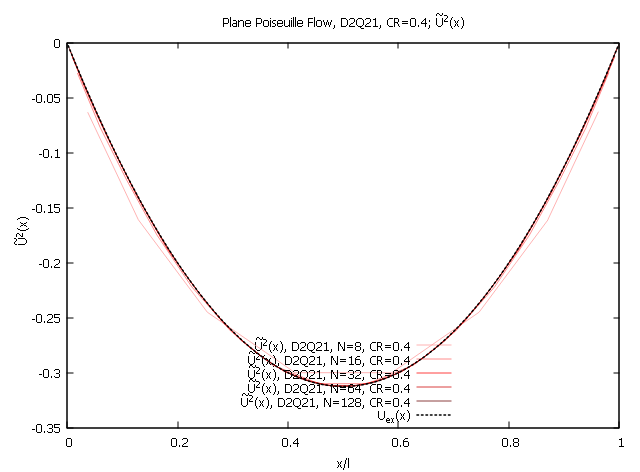}
\end{subfigure}
\end{figure}

In Figs. \ref{PPD2Q9Rho}-\ref{PPD2Q21Ut2} below we present the numerical solution to the non-equidistant case with compression of cells with \(CR=0.4\). The results for \(\rho(x)\) and \(\tilde{U}^2(x)\),  for D2Q9 and \(CR=0.4\) are shown in Figs. \ref{PPD2Q9Rho},\ref{PPD2Q9Ut2} below. 
\(\tilde{U}^1\) in this case is negligibly small for all resolutions. We again observe a very good convergence with the increase in resolution to the exact solution Eqs. \mref{PlanePoiseuilleSolution2,PlanePoiseuilleSolutionLatticeUnits}. The results for \(\rho(x)\) and  \(\tilde{U}^2\) for D2Q21 and \(CR=0.4\) are shown in Figs. \ref{PPD2Q21Rho}, \ref{PPD2Q21Ut2} below. 
We again observe a very good convergence with the increase in resolution to the exact solution Eqs. \mref{PlanePoiseuilleSolution2,PlanePoiseuilleSolutionLatticeUnits}. One can notice a better performance of the D2Q21 lattice in the middle of the domain than that of the D2Q9. 

Let us now show the results of numerical model solutions for a fixed resolution \(N=64\) but variable \(\tau=\{1.0,0.9,0.8,0.7\}\) for both lattices D2Q9 and D2Q21. 
The corresponding LBM parameters for the D2Q9 lattice we used are shown in \Cref{PPTau1}, \Cref{PPD2Q9N64} above. 

The agreement of numerical with the exact solution with no lattice compression \(CR=0\) was very good for both lattices D2Q9 and D2Q21 for all \(\tau=\{1.0,0.9,0.8,0.7\}\). Below in Figs. \ref{PPD2Q9CRUt2Ulat}, \ref{PPD2Q9CRUt2Uex} we present the comparisons of numerical results with the exact solution for lattice D2Q9 with compression \(CR=0.35\) and \(\tau=\{1.0,0.9,0.8,0.7\}\). We show variable \(\tau\) cases on the same graphs, for the following quantities: \(\frac{\overline{U}^2(x)}{U_{lat}}, \frac{U_{ex}(x)}{U_{lat}}\), and  \(\frac{\overline{U}^2(x)}{\overline{U}^2_{ex}(x)}\). 
\begin{figure}[H]
\centering
\caption{Plane Poiseuille flow, contracting grid, variable \(\tau\) for D2Q9 lattice}
\begin{subfigure}[H]{0.6\textwidth}
\centering
\subcaption{\(\frac{\overline{U}^2(x)}{U_{lat}}, \frac{U_{ex}(x)}{U_{lat}}\) for D2Q9, \(\tau=1.0,0.9,0.8,0.7\),  and \(CR=0.35\)}
\label{PPD2Q9CRUt2Ulat}
\includegraphics[width=\textwidth]{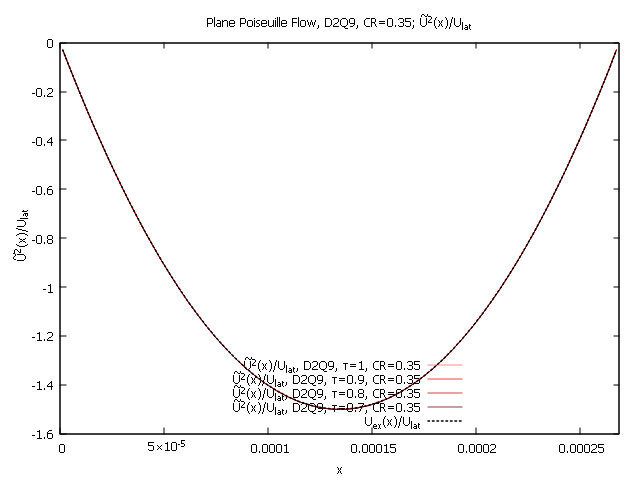}
\end{subfigure}
\begin{subfigure}[H]{0.6\textwidth}
\centering
\caption{\(\frac{\overline{U}^2(x)}{\overline{U}^2_{ex}(x)}\) for D2Q9, \(\tau=1.0,0.9,0.8,0.7\),  and \(CR=0.35\)}
\label{PPD2Q9CRUt2Uex}
\includegraphics[width=\textwidth]{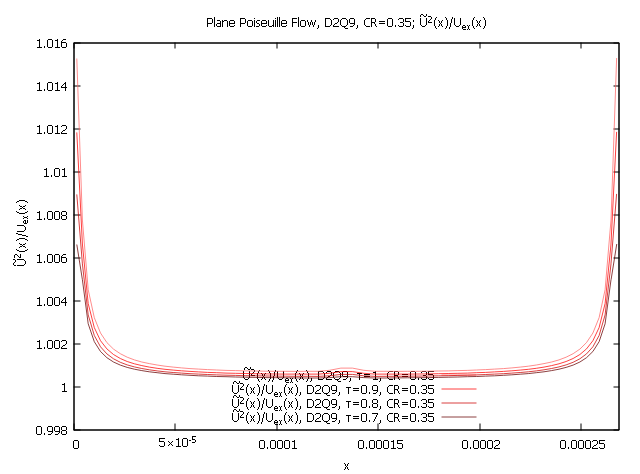}
\end{subfigure}
\end{figure}
The \(\frac{\overline{U}^1(x)}{U_{lat}}\) for D2Q9, \(\tau=1.0,0.9,0.8,0.7\),  and \(CR=0.35\), are all equal to zero within double-precision accuracy and the \(\frac{\rho(x)}{\rho_0}\) for D2Q9, \(\tau=1.0,0.9,0.8,0.7\),  and \(CR=0.35\) show very weak dependence on \(\tau\). As one can see, the new method \cite{HChen} using the D2Q9 lattice with strong cell compression works well for various values of parameter \(\tau\). 

\begin{figure}[H]
\centering
\caption{Plane Poiseuille flow, contracting grid, variable \(\tau\) for D2Q21 lattice}
\begin{subfigure}[H]{0.6\textwidth}
\centering
\subcaption{\(\frac{\overline{U}^2(x)}{U_{lat}}, \frac{U_{ex}(x)}{U_{lat}}\) for D2Q21, \(\tau=1.0,0.9,0.8,0.7\),  and \(CR=0.35\)}
\label{PPD2Q21CRUt2Ulat}
\includegraphics[width=\textwidth]{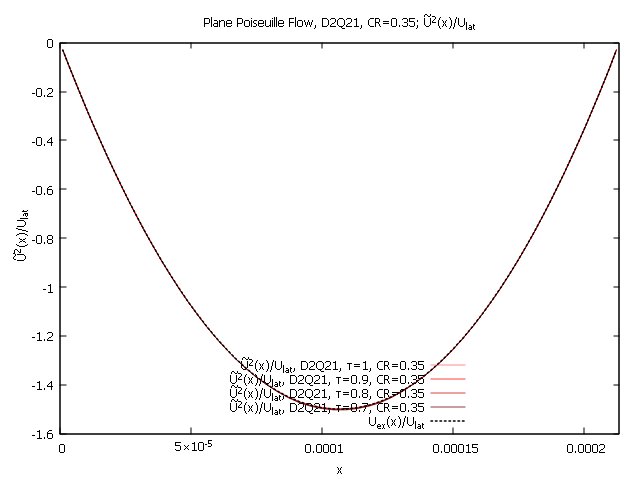}
\end{subfigure}
\begin{subfigure}[H]{0.6\textwidth}
\centering
\subcaption{\(\frac{\overline{U}^2(x)}{\overline{U}^2_{ex}(x)}\) for D2Q21, \(\tau=1.0,0.9,0.8,0.7\),  and \(CR=0.35\)}
\label{PPD2Q21CRUt2Uex}
\includegraphics[width=\textwidth]{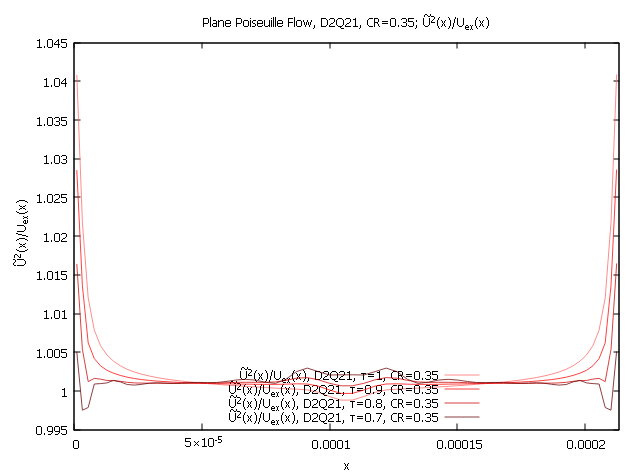}
\end{subfigure}
\end{figure}

The corresponding LBM parameters for the D2Q21 lattice we used are shown in \Cref{PPTau1}, \Cref{PPD2Q21N64} above. In Figs. \ref{PPD2Q21CRUt2Ulat}, \ref{PPD2Q21CRUt2Uex} below we present the numerical results for D2Q21 lattice. 
Similar to the D2Q9 case, the \(\frac{\overline{U}^1(x)}{U_{lat}}\) for D2Q21, \(\tau=1.0,0.9,0.8,0.7\),  and \(CR=0.35\), are approximately equal to zero and the \(\frac{\rho(x)}{\rho_0}\) for D2Q21, \(\tau=1.0,0.9,0.8,0.7\),  and \(CR=0.35\) show very weak dependence on \(\tau\). Here we observe that the new method \cite{HChen} with strong compression also works well for various values of parameter \(\tau\). 

To summarize, for these two planar geometry problems above, both D2Q9 and D2Q21 lattices provide adequate and comparable performance for cases with and without lattice compression. Let us now move on to problems with intrinsic curvilinearity. 

\subsection{Circular Couette Flow}

This is also a well-known \cite{Batchelor} naturally curvilinear problem with a closed-form exact solution, which fits very well for our purpose of validating the new method. This solution describes a stationary flow of viscous liquid between two infinite concentric vertical cylinders with radii \(R_1\) for the internal one and \(R_2\) for the external one, rotating with corresponding angular velocities \(\Omega_1\) and \(\Omega_2\). Here we will need the leading order in \(Ma=\frac{U_{avg}}{c_s}\) (Mach number) exact solution of the compressible Navier-Stokes equations. 

For both of the circular problems considered here, we can introduce a measure of natural curvilinearity, a coefficient \(NCR=\frac{R_2}{R_1}\). If we keep the distance between cylinders \(R_2-R_1=const\), then \(NCR=1\) corresponds to a previously considered planar case for \(R_1\rightarrow +\infty\). Conversely, cases \(NCR\gg 1\) correspond to cases with strong natural curvilinearity. The Reynolds number for this problem was defined as \(Re=\frac{\Omega_1 R_1\left(R_2-R_1\right)}{\nu}\). An example of geometry and the lattice  with strong natural curvilinearity with \(NCR=11\) that we have used in the calculations is shown in Fig. \ref{CircularGeometryNCR11}. 

The \(\theta\)-component of the Navier-Stokes equations written in polar coordinates \(\vec{r}=\{r,\theta\}\) in our case of a stationary \(r\)-dependent, \(\theta\)-directional flow is: 
\[
-\frac{u_{\theta}}{r^2}+\frac{1}{r}\frac{d}{d r}\left(r\frac{d u_{\theta}}{d r}\right)=0,
\]
with the following solution: 
\begin{equation}
\label{CircularCouetteSolution1}
u_{\theta}\left(r\right)=ar+\frac{b}{r}; 
\end{equation}

\begin{equation}
\label{CircularCouetteSolution2}
a=\frac{\Omega_2R_2^2-\Omega_1R_1^2}{R_2^2-R_1^2}; 
b=\frac{R_1^2R_2^2\left(\Omega_1-\Omega_2\right)}{R_2^2-R_1^2}. 
\end{equation}
Note that this classical solution has the following properties: it is a non-monotonic function of \(r\) for \(0\leq\frac{\Omega_2}{\Omega_1}<1\) with an extremum at 
\[
\hat{r}=\sqrt{\frac{\Omega_2-\Omega_1}{\frac{\Omega_1}{R_2^2}-\frac{\Omega_2}{R_1^2}}},
\] 
and it is a monotone function outside of this interval. 

Also, note that the above velocity profile has the same leading order behavior in small parameters \(0\leq\frac{x}{R_1}\leq\frac{l}{R_1}\) for \(l=R_2-R_1\) and \(x=r-R_1\) as the exact planar Couette flow solution when both vertical planes are moving with velocities \(U_1=\Omega_1R_1\) and \(U_2=\Omega_2R_2\): 
\[
u_{\theta}(x)\sim U_1+\left(U_2-U_1\right)x/l, 
\]
which for \(U_1=0\) reproduces the exact solution for the plane Couette flow Eq.  \ref{PlaneCouetteExactSolution}. 

The radial component of the Navier-Stokes equation for a stationary \(r\)-dependent flow only in \(\theta\)-direction, is:
\begin{equation}
\label{RadialComponent}
-\frac{u_{\theta}^2}{r}=-\frac{1}{\rho_0}\frac{\partial p}{\partial r}, 
\end{equation}
which simplifies into the following \(1\)-st order ODE: 
\[
\frac{1}{\rho_0}\frac{dp(r)}{dr}=\frac{1}{r}\left(ar+\frac{b}{r}\right)^2.
\]
The LBM formalism results in an expansion in powers of \(Ma\) number with the ideal gas law equation of state: 
\begin{equation}
\label{EquationOfState}
p(r)=\rho(r)T_0.
\end{equation}
If we substitute this into the previous incompressible equation, we obtain the leading order in \(Ma\) ODE for density: \[
\frac{1}{\rho_0}\frac{d\rho(r)}{dr}=\frac{1}{rT_0}\left(ar+\frac{b}{r}\right)^2.
\]
This defines the density behavior at the leading order in \(Ma\): 
\[
\rho(r)=\tilde{\rho}_0+\frac{\rho_0}{T_0}h(r),
\]
where: 
\[
h(r)=-\frac{b^2}{2r^2}+2ab\log\left(\frac{r}{R_1}\right)+\frac{a^2r^2}{2}.
\]
The integration constant \(\tilde{\rho}_0\) can be determined using conservation of total mass: 
\[
M_0=\int_{R_1}^{R_2}r\rho(r)dr=\rho_0\pi\left(R_2^2-R_2^2\right), 
\]
which leads to: 
\[
\tilde{\rho}_0=\rho_0\left(1-\frac{2H(R_1,R_2)}{T_0\left(R_2^2-R_1^2\right)}\right), 
\]
where we denoted: 
\begin{equation}
\label{CircularCouetteSolution3}
H(R_1,R_2)=-\frac{b^2}{2}\log\left(\frac{R_2}{R_1}\right)+ab\left(R_2^2\log\left(\frac{R_2}{R_1}\right)-\frac{R_2^2-R_1^2}{2}\right)+\frac{a^2}{8}\left(R_2^4-R_1^4\right).
\end{equation}
Thus the leading-order in \(Ma\) exact solution for density is: 
\begin{equation}
\label{CircularCouetteSolution4}
\rho(r)=\rho_0\bigg\{1+\frac{1}{T_0}\left(h(r)-\frac{2H\left(R_1,R_2\right)}{R_2^2-R_1^2}\right)\bigg\}.
\end{equation}

We will need the above exact solution for velocity expressed in lattice units on the non-equidistant lattice: 
\begin{equation}
\label{CircularCouetteSolutionLatticeUnits1}
u^{lat}_{{\theta},i}=a_{lat}\frac{r_i}{\overline{\triangle}_r}+\frac{b_{lat}\overline{\triangle}_r}{r_i},
\end{equation}
where we denoted: 
\begin{equation}
\label{CircularCouetteSolutionLatticeUnits2}
a_{lat}=\frac{\omega^{lat}_2\left(M+N_r\right)^2-\omega^{lat}_1M^2}{N_r\left(2M+N_r\right)} \mbox{ and } b_{lat}=\frac{M^2\left(N_r+M\right)^2\left(\omega^{lat}_1-\omega^{lat}_2\right)}{N_r\left(2M+N_r\right)},
\end{equation}
and similar to earlier definitions,  \(\overline{\triangle}_r=\frac{R_2-R_1}{N_r}\) is the average step in the radial direction. 
Similarly, the exact solution for density in lattice units on the non-equidistant lattice is: \begin{equation}
\label{CircularCouetteSolutionLatticeUnits3}
\rho^{lat}_i=\rho^{lat}_0\bigg\{1+\frac{1}{T^{lat}_0}\left(h^{lat}_i-\frac{H^{lat}\left(M,N_r\right)}{N_r\left(M+\frac{N_r}{2}\right)}\right)\bigg\},
\end{equation}
where we denoted: 
\begin{equation}
\label{CircularCouetteSolutionLatticeUnits4}
h_i^{lat}=-\frac{b_{lat}^2\overline{\triangle}_r^2}{2r_i^2}+2a_{lat}b_{lat}\log\left(\frac{r_i}{\overline{\triangle}_rM}\right)+\frac{a_{lat}^2r_i^2}{2\overline{\triangle}_r^2},
\end{equation}
and
\begin{equation}
\label{CircularCouetteSolutionLatticeUnits5}
\begin{split}
H^{lat}\left(M,N_r\right)=&-\frac{b_{lat}^2}{2}\log\left(1+\frac{N_r}{M}\right)+a_{lat}b_{lat}\bigg\{\left(M+N_r\right)^2\log\left(1+\frac{N_r}{M}\right)- \\
&-N_r\left(M+\frac{N_r}{2}\right)\bigg\}+\frac{a_{lat}^2}{8}\bigg\{\left(M+N_r\right)^4-M^4\bigg\}.
\end{split}
\end{equation}

In both of our circular problems, we used the same method of choosing the azimuthal lattice size \(N_{\theta}\) as a function of radial lattice size \(N_r\), which was the following. Requiring that the lattice cells near \(r=R_1\) are approximately equilateral results in \(N_{\theta}=N_r\frac{2\pi R_1}{R_2-R_1}\). 

Obviously, such intrinsically curvilinear problems cannot be solved by a standard LBM even without radial lattice step contraction. Therefore, here we present the comparisons between D2Q9 and D2Q21 lattices without radial grid contraction for a case of strong natural curvilinearity \(NCR=11\). We have considered the case of \(\Omega_2=0\) in the numerical solutions below. 

The LBM parameters for the D2Q9 lattice we have used are presented in \Cref{TableCCTau1}, \Cref{CPD2Q9}, and the LBM parameters used for the D2Q21 lattice are presented in \Cref{TableCCTau1}, \Cref{CPD2Q21}. 
\begin{table}[H]
\centering
\caption{Circular Couette LBM parameters}
\label{TableCCTau1}
\begin{subtable}{\hsize}
\centering
\subcaption{D2Q9 lattice, variable \(N_x\), \(\tau=1\)}
\label{CPD2Q9}
\scalebox{0.93}{
\begin{tabular}{|p{0.01cm}p{0.3cm}|p{1.7cm}p{0.01cm}p{0.01cm}p{0.01cm}p{0.4cm}|}
\hline
\(N_{\theta}\) & \(N_r\) & \(\Omega_{1,lat}\) & \(\nu_{lat}\) & \(U_{lat}\) & \(Ma_{sim}\) & \(Re\) \\
\hline
\(40\) & \(64\) & \(3.82\times 10^{-2}\) & \(0.167\) & \(0.245\) & \(0.424\) & \(93.9\) \\
\(80\) & \(128\) & \(1.91\times 10^{-2}\) & \(0.167\) & \(0.245\) & \(0.424\) & \(187.9\) \\
\(160\) & \(256\) & \(9.56\times 10^{-3}\) & \(0.167\) & \(0.245\) & \(0.424\) & \(375.8\) \\
\(320\) & \(512\) & \(4.78\times 10^{-3}\) & \(0.167\) & \(0.245\) & \(0.424\) & \(751.5\) \\
\(640\) & \(1024\) & \(2.39\times 10^{-3}\) & \(0.167\) & \(0.245\) & \(0.424\) & \(1503.1\) \\
\hline
\end{tabular}}
\end{subtable}
\begin{subtable}{\hsize}
\centering
\subcaption{D2Q21 lattice, variable \(N_x\), \(\tau=1\)}
\label{CPD2Q21}
\centering
\scalebox{0.93}{
\begin{tabular}{|p{0.01cm}p{0.3cm}|p{1.7cm}p{0.01cm}p{0.01cm}p{0.01cm}p{0.4cm}|}
\hline
\(N_{\theta}\) & \(N_r\) & \(\Omega_{1,lat}\) & \(\nu_{lat}\) & \(U_{lat}\) & \(Ma_{sim}\) & \(Re\) \\
\hline
\(40\) & \(64\) & \(3.82\times 10^{-2}\) & \(0.333\) & \(0.245\) & \(0.300\) & \(47.0\) \\
\(80\) & \(128\) & \(1.91\times 10^{-2}\) & \(0.333\) & \(0.245\) & \(0.300\) & \(93.9\) \\
\(160\) & \(256\) & \(9.56\times 10^{-3}\) & \(0.333\) & \(0.245\) & \(0.300\) & \(187.9\) \\
\(320\) & \(512\) & \(4.78\times 10^{-3}\) & \(0.333\) & \(0.245\) & \(0.300\) & \(375.8\) \\
\(640\) & \(1024\) & \(2.39\times 10^{-3}\) & \(0.333\) & \(0.245\) & \(0.300\) & \(751.5\) \\
\hline
\end{tabular}}
\end{subtable}
\end{table}

\begin{figure}[H]
\centering
\caption{Circular Couette flow, \(NCR=11\), variable \(N_r\), D2Q9 lattice}
\label{CCD2Q9}
\begin{subfigure}[H]{0.51\textwidth}
\centering
\subcaption{\(\rho(r)\) for \(N_r=64,128,256,512,1024\) and \(NCR=11\)}
\label{CCD2Q9Rho}
\includegraphics[width=\textwidth]{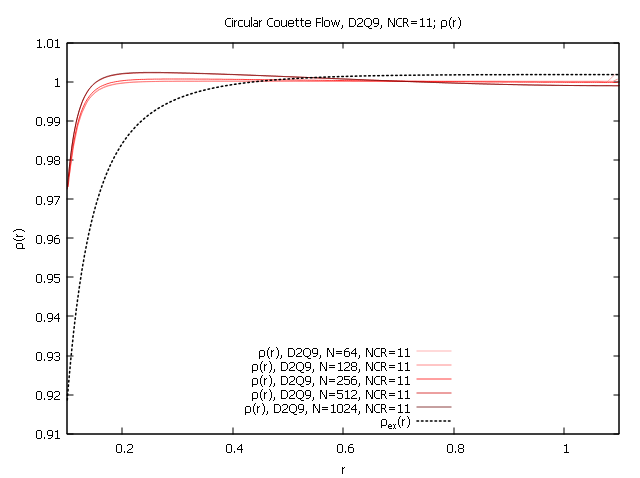}
\end{subfigure}
\begin{subfigure}[H]{0.51\textwidth}
\centering
\subcaption{\(\tilde{U}^r(r)\) for \(N_r=64,128,256,512,1024\) and \(NCR=11\)}
\label{CCD2Q9Utr}
\includegraphics[width=\textwidth]{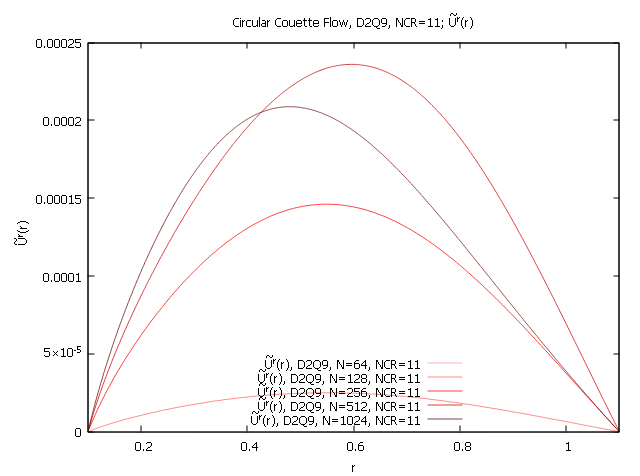}
\end{subfigure}
\begin{subfigure}[H]{0.51\textwidth}
\centering
\subcaption{\(\tilde{U}^{\theta}(r)\) for \(N_r=64,128,256,512,1024\) and \(NCR=11\)}
\label{CCD2Q9Uttheta}
\includegraphics[width=\textwidth]{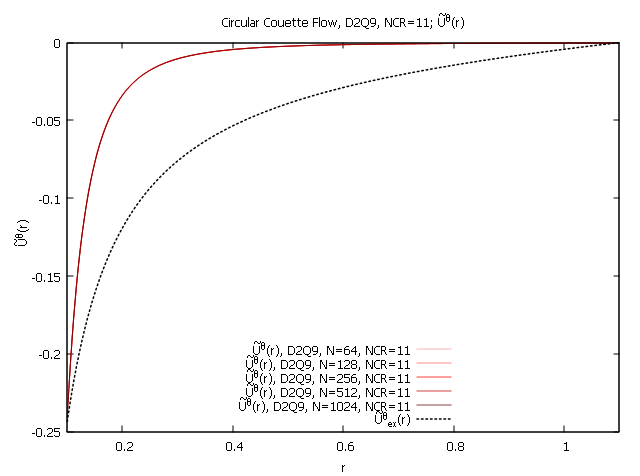}
\end{subfigure}
\end{figure}

In \Cref{CCD2Q9Rho}-\Cref{CCD2Q9Uttheta} we present the numerical solutions for \(\rho(r), \tilde{U}^r(r)\) and \(\tilde{U}^{\theta}(r)\) for resolutions \(N_r=64,128,256,512\), and \(1024\) for D2Q9 lattice. Note that in the cases below we only show \(\rho\) after the application of the "no flow" adjustment described above.
In a drastic difference to the previously considered planar problems, we observe here that the D2Q9 lattice does not converge to the exact solutions Eqs.  \mref{CircularCouetteSolution1,CircularCouetteSolution4,CircularCouetteSolutionLatticeUnits1,CircularCouetteSolutionLatticeUnits3}. This calls for exploring of the higher-order D2Q21 lattice. 

\begin{figure}[H]
\centering
\caption{Circular Couette flow, \(NCR=11\), variable \(N_r\), D2Q21 lattice}
\begin{subfigure}[H]{0.51\textwidth}
\centering
\caption{\(\rho(r)\) for \(N_r=64,128,256,512,1024\) and \(NCR=11\)}
\label{CCD2Q21Rho}
\includegraphics[width=\textwidth]{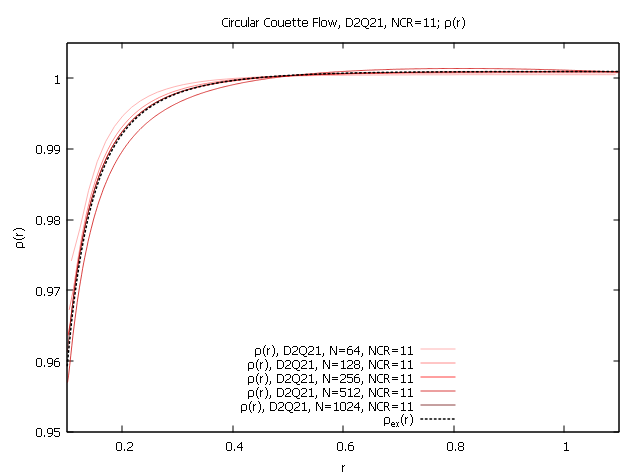}
\end{subfigure}
\begin{subfigure}[H]{0.51\textwidth}
\centering
\caption{\(\tilde{U}^r(r)\) for \(N_r=64,128,256,512,1024\) and \(NCR=11\)}
\label{CCD2Q21Utr}
\includegraphics[width=\textwidth]{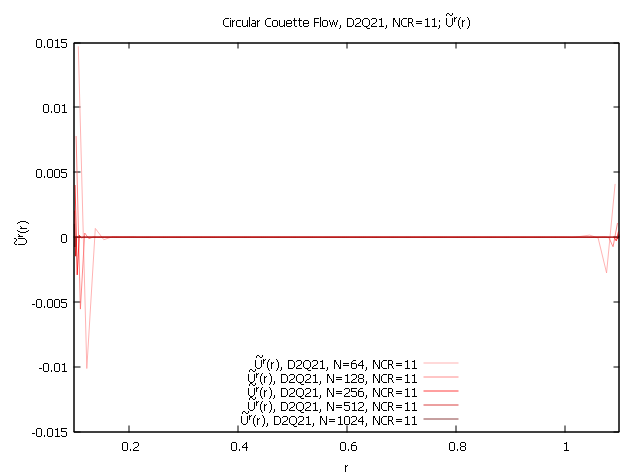}
\end{subfigure}
\begin{subfigure}[H]{0.51\textwidth}
\centering
\caption{\(\tilde{U}^{\theta}(r)\) for \(N_r=64,128,256,512,1024\), and \(NCR=11\)}
\label{CCD2Q21Uttheta}
\includegraphics[width=\textwidth]{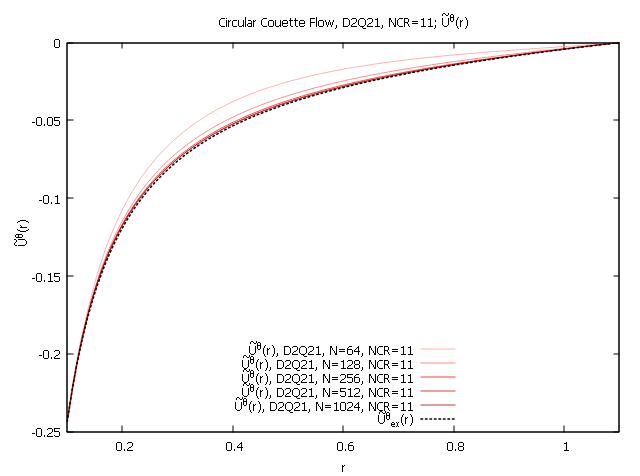}
\end{subfigure}
\end{figure}

\Cref{CCD2Q21Rho}-\Cref{CCD2Q21Uttheta} present the numerical solutions for \(\rho(r), \tilde{U}^r(r)\) and \(\tilde{U}^{\theta}(r)\) for resolutions \(N_r=64,128,256,512\), and \(1024\) for D2Q21 lattice, which possesses a higher degree of isotropy (see Appendix A). 
Notice that unlike in the D2Q9 case, the implementation of the model based on the D2Q21 lattice does converge very well to the exact solutions Eqs.  \mref{CircularCouetteSolution1,CircularCouetteSolution4,CircularCouetteSolutionLatticeUnits1,CircularCouetteSolutionLatticeUnits3} with the increase in resolution. The more widely used D2Q9 lattice simply does not have sufficient isotropy to support the LBM on the curvilinear mesh. 

Let us try to outline here the reasons for the importance of higher-order isotropy. As shown in \cite{HChen}, the 6th-order isotropy is required to reproduce the Navier-Stokes equation in curvilinear coordinates. In particular, the 3-rd order tensor \(Q^{ijk,eq}\) will satisfy the 6th order isotropy condition given by the Eq. (32) of \cite{HChen} for a specific choice of equilibrium distribution function \(f^{eq}_{\alpha}\), which requires the Hermite expansion up to 3rd-order, given by the Eq. (33) of \cite{HChen}. Therefore, the 6th-order isotropy as defined in the last Eq. (10) of \cite{HChen} is required in order to avoid discrete rotational artifacts. 

\subsection{Circular Flow Driven By an Azimuthal Force (Circular Poiseuille Flow)}

Similar to the previous case, this is a naturally curvilinear problem with the same exact geometry, but it has a closed-form exact solution which in the limiting case of \( \frac{R_2-R_1}{R_1}\rightarrow 0+\), such that \(R_2-R_1=const\) converges to the Plane Poiseuille flow considered above. For this problem, in addition to the Reynolds number \(Re=\frac{\overline{U}\left(R_2-R_1\right)}{\nu}\), we can also define the Froude number as \(Fr=\frac{\overline{U}}{\sqrt{g\left(R_2-R_1\right)}}\), with the average velocity \(\overline{U}\) obtained from the exact solution below. 

This is a model problem that we at the present time can not connect to any physical situation but is useful for analyses of numerical aspects associated with curvilinear LBM. We are not aware that this problem was considered before. 

As is shown in Appendix D, a constant curvilinear second co-tangent component of the external "gravity" force \(g^2\) corresponds to the following external "gravity" component in polar coordinates: 
\begin{equation}
g_{\theta}=g^2\frac{\sin\triangle_{\theta}}{\overline{\triangle}_r}r_i.
\end{equation}
Denote \(\alpha=\frac{\sin\triangle_{\theta}}{\overline{\triangle}_r}\), then the \(\theta\)-component of the  external "gravity" force corresponds to the  acceleration \(g_{\theta}=\alpha rg^2\). 
Similarly to the previous section, in the particular case of stationary \(r\)-dependent and \(\theta\)-directional flow in the presence of external force with acceleration 
\(\vec{g}=\begin{pmatrix} 0 \\ -\alpha rg \end{pmatrix}\), which is pushing the fluid in the clockwise azimuthal direction for \(\alpha>0\), we get: 
\begin{equation}
\frac{\mu}{\rho_0}\bigg\{-\frac{u_{\theta}}{r^2}+\frac{1}{r}\frac{d}{dr}\left(r\frac{du_{\theta}}{dr}\right)\bigg\}+\alpha rg=0,
\end{equation}
resulting in: 
\begin{equation}
\label{CircularPoiseuilleSolution1}
u_{\theta}(r)=ar+\frac{b}{r}+\frac{A}{8}r^3.
\end{equation}
Here \(A=\alpha\frac{g}{\nu}\), \(\nu=\frac{\mu}{\rho_0}\) is the  kinematic viscosity, and the constants \(a\) and \(b\) determined from the no-slip boundary conditions \(u_{\theta}\left(R_1\right)=u_{\theta}\left(R_2\right)=0\): 
\begin{equation}
a=-\frac{A}{8}\left(R_1^2+R_2^2\right)<0, \mbox{ and } b=\frac{A}{8}R_1^2R_2^2>0.
\end{equation}
The velocity profile given by the Eq. \eqref{CircularPoiseuilleSolution1} which starts and ends at \(0\) and has a minimum at: 
\begin{equation}
\hat{r}=\sqrt{\frac{4}{3A}\left(-a+\sqrt{a^2+\frac{3Ab}{2}}\right)}.
\end{equation}
One can also define the average velocity which can be used as a characteristic flow velocity for the specification of Reynolds and Froude numbers: 
\begin{equation}
\overline{u}_{\theta}=\frac{1}{R_2-R_1}\int_{R_1}^{R_2}u_{\theta}(r)dr=\frac{a}{2}\left(R_2+R_1\right)+b\frac{\log{\frac{R_2}{R_1}}}{R_2-R_1}+\frac{A}{32}\left(R_2+R_1\right)\left(R_2^2+R_1^2\right).
\end{equation}
Note that the leading-order behavior of this curvilinear flow in the small parameters \(x\), \(\epsilon\): \(0\leq x=\frac{r-R_1}{R_1}\leq \epsilon=\frac{R_2-R_1}{R_1}\), is: 
\[
\frac{u_{\theta}(r)}{\overline{\triangle}_r\alpha R_1^3}\sim\frac{A_0}{2}x\left(x-\epsilon\right),
\]
where we denoted \(A_0=\frac{g}{\nu}\), which exactly corresponds to the plane Poiseuille flow in Cartesian coordinates Eq. \ref{PlanePoiseuilleSolution2}. 

The radial component of the Navier-Stokes equation Eq. \eqref{RadialComponent}
with the equation of state given by the Eq. \eqref{EquationOfState}
becomes
\begin{equation}
\frac{1}{\rho_0}\frac{d\rho(r)}{dr}=\frac{1}{rT_0}\left(ar+\frac{b}{r}+\frac{A}{8}r^3\right)^2, 
\end{equation}
with the solution: 
\begin{equation}
\rho(r)=\rho_0\bigg[1+\frac{1}{T_0}\left(h(r)-\frac{2H\left(R_1,R_2\right)}{R_2^2-R_1^2}\right)\bigg],
\end{equation}
where: 
\begin{equation}
h(r)=-\frac{b^2}{2r^2}+2ab\log\frac{r}{R_1}+\frac{1}{2}\left(a^2+\frac{bA}{4}\right)r^2+\frac{aAr^4}{16}+\frac{A^2r^6}{384},
\end{equation}
and
\begin{equation}
\begin{split}
H(R_1,R_2)=&-\frac{b^2}{2}\log\frac{R_2}{R_1}+ab\left(R_2^2\log\frac{R_2}{R_1}-\frac{R_2^2-R_1^2}{2}\right)+ \\
&+\frac{1}{8}\left(a^2+\frac{bA}{4}\right)\left(R_2^4-R_1^4\right)+\frac{aA}{96}\left(R_2^6-R_1^6\right)+\frac{A^2}{3072}\left(R_2^8-R_1^8\right).
\end{split}
\end{equation}
Rewritten in lattice units, 
\begin{equation}
\label{CircularPoiseuilleSolutionLatticeUnits1}
u_{\theta,i}^{lat}=a_{lat}\frac{r_i}{\overline{\triangle}_r}+b_{lat}\frac{\overline{\triangle}_r}{r_i}+\frac{A_{lat}}{8}\frac{r_i^3}{\overline{\triangle}_r^3},
\end{equation}
\begin{equation}
A_{lat}=\frac{g_{lat}}{\nu_{lat}}, \mbox{ } a_{lat}=-\frac{A_{lat}}{8}\bigg[\left(N_r+M\right)^2+M^2\bigg], \mbox{ } b_{lat}=\frac{A_{lat}}{8}\left(N_r+M\right)^2; 
\end{equation}
\begin{equation}
\label{CircularPoiseuilleSolutionLatticeUnits2}
\rho_{\theta,i}^{lat}=\rho_0^{lat}\bigg\{1+\frac{1}{T_0^{lat}}\left(h_i^{lat}-\frac{H^{lat}\left(M,N_r\right)}{N_r\left(M+\frac{N_r}{2}\right)}\right)\bigg\}, 
\end{equation}
\begin{equation}
\begin{split}
h^{lat}_i=&-\frac{b_{lat}^2\overline{\triangle}_r^2}{2r_i^2}+2a_{lat}b_{lat}\log\frac{r_i}{\overline{\triangle}_rM}+\frac{1}{2}\left(a_{lat}^2+\frac{b_{lat}A_{lat}}{4}\right)\frac{r_i^2}{\overline{\triangle}_r^2}+ \\
&+\frac{a_{lat}A_{lat}}{16}\frac{r_i^4}{\overline{\triangle}_r^4}+\frac{A_{lat}^2}{384}\frac{r_i^6}{\overline{\triangle}_r^6},
\end{split}
\end{equation}
\begin{equation}
\begin{split}
H^{lat}\left(M,N_r\right)=&-\frac{b_{lat}^2}{2}\log\left(1+\frac{N_r}{M}\right)+a_{lat}b_{lat}\bigg[\left(M+N_r\right)^2\log\left(1+\frac{N_r}{M}\right)- \\
&-N_r\left(m+\frac{N_r}{2}\right)\bigg]+\frac{1}{8}\left(a_{lat}^2+\frac{b_{lat}A_{lat}}{4}\right)\bigg[\left(M+N_r\right)^4-M^4\bigg]+ \\
&+\frac{a_{lat}A_{lat}}{96}\bigg[\left(M+N_r\right)^6-M^6\bigg]+\frac{A_{lat}^2}{3072}\bigg[\left(M+N_r\right)^8-M^8\bigg].
\end{split}
\end{equation}

As we have shown in the previous section, on non-equilateral grid systems the D2Q9 lattice cannot adequately reproduce the exact solutions in intrinsically curvilinear geometries due to the lack of \(6\)-th order isotropy. Therefore we will be only showing here the results for the D2Q21 lattice. We have again used the same case of strong intrinsic curvilinearity \(NCR=11\). 

\begin{figure}[H]
\centering
\caption{Circular Poiseuille flow, \(NCR=11\), variable \(N_r\), D2Q21 lattice}
\begin{subfigure}[H]{0.51\textwidth}
\centering
\subcaption{\(\rho(r)\) for \(N_r=64,128,256,512,1024\) and \(NCR=11\)}
\label{CPD2Q21Rho}
\includegraphics[width=\textwidth]{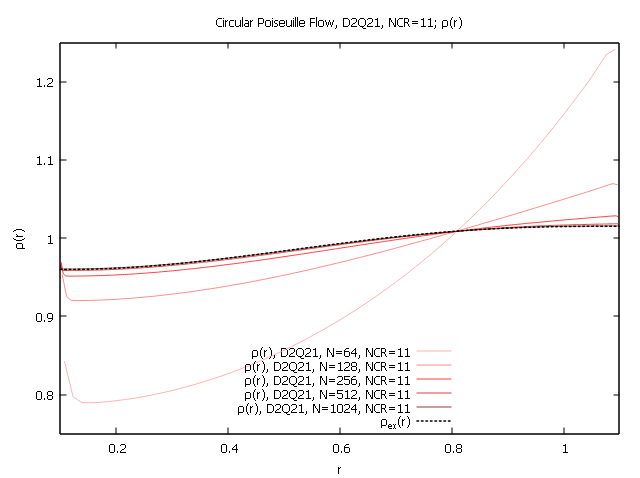}
\end{subfigure}
\begin{subfigure}[H]{0.51\textwidth}
\centering
\subcaption{\(\tilde{U}^r(r)\) for \(N_r=64,128,256,512,1024\) and \(NCR=11\)}
\label{CPD2Q21Utr}
\includegraphics[width=\textwidth]{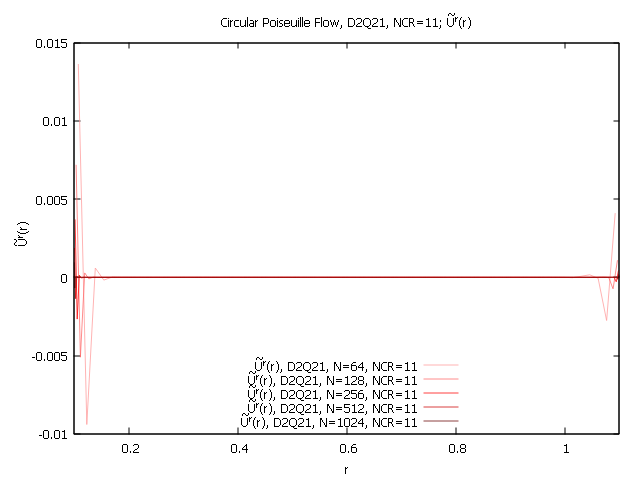}
\end{subfigure}
\begin{subfigure}[H]{0.51\textwidth}
\centering
\subcaption{\(\tilde{U}^{\theta}(r)\) for \(N_r=64,128,256,512,1024\) and \(NCR=11\)}
\label{CPD2Q21Uttheta}
\includegraphics[width=\textwidth]{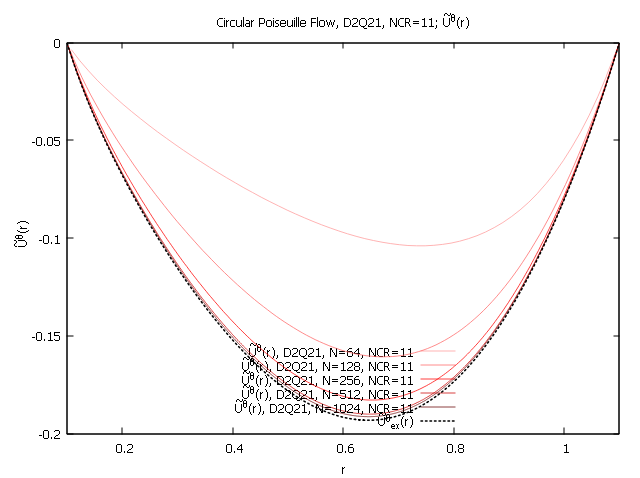}
\end{subfigure}
\end{figure}

The Tbl. \ref{TableCPD2Q21Tau1} below lists the set of LBM parameters used for comparisons in this case: 
\begin{table}[H]
\centering
\scalebox{0.93}{
\begin{tabular}{|p{0.01cm}p{0.3cm}|p{0.1cm}p{0.01cm}p{1.65cm}p{0.01cm}p{0.1cm}p{0.4cm}|}
\hline
\multicolumn{8}{|c|}{Circular Poiseuille, D2Q21 Lattice, \(\tau=1\)} \\
\hline
\(N_{\theta}\) & \(N_r\) & \(\nu_{lat}\) & \(U_{lat}\) & \(g_{lat}\) & \(Ma_{sim}\) & \(Re\) & \(Fr\) \\
\hline
\(40\) & \(64\) & \(0.334\) & \(0.130\) & \(2.52\times 10^{-5}\) & \(0.159\) & \(24.9\) & \(3.2\) \\
\(80\) & \(128\) & \(0.333\) & \(0.130\) & \(6.26\times 10^{-6}\) & \(0.159\) & \(50.0\) & \(4.6\) \\
\(160\) & \(256\) & \(0.333\) & \(0.130\) & \(1.56\times 10^{-6}\) & \(0.159\) & \(100.1\) & \(6.5\) \\
\(322\) & \(512\) & \(0.335\) & \(0.130\) & \(3.96\times 10^{-7}\) & \(0.159\) & \(199.0\) & \(9.1\) \\
\(644\) & \(1024\) & \(0.335\) & \(0.130\) & \(9.89\times 10^{-8}\) & \(0.159\) & \(397.9\) & \(12.9\) \\
\hline
\end{tabular}}
\caption{Circular Poiseuille flow, D2Q21 lattice parameters, \(\tau=1\)}
\label{TableCPD2Q21Tau1}
\end{table}
The numerical solutions for \(\rho(r), \tilde{U}^r(r)\) and \(\tilde{U}^{\theta}(r)\) for resolutions \(N_r=64,128,256,512\), and \(1024\) for D2Q21 lattice are presented in Figs. \ref{CPD2Q21Rho}-\ref{CPD2Q21Uttheta}. 
We again observe that the new fully curvilinear method \cite{HChen} using D2Q21 converges very well to the exact solutions Eqs.  \mref{CircularPoiseuilleSolution1,RadialComponent,CircularPoiseuilleSolutionLatticeUnits1,CircularPoiseuilleSolutionLatticeUnits2} with the increase in resolution. 

A somewhat high number of grid points (such as \(N_r=512\) and higher for the Circular Poiseuille problem considered above) was required to get sufficient accuracy for the curvilinear cases. We do not see this however as a major obstacle to the practical implementation of the new method. The observed high-resolution requirement is partially due to some imperfections in the boundary condition algorithm. This paper is the first to numerically establish that the current curvilinear LBM is able to recover the Navier-Stokes hydrodynamics asymptotically. How to improve the rate of convergence with a lower number of grid points is a research topic for the future. 

\section{Discussion}

In this work, we have provided the first numerical tests of the novel volumetric curvilinear LBM method \cite{HChen}. The accuracy and performance of the new method have been investigated and validated on a set of four 2D exactly solvable models: with and without the natural curvilinearity. The crucial importance of the isotropy requirement of at least 6th-order which requires at least D2Q21 lattice in 2D cases has also been demonstrated. We considered four 2-dimensional exactly solvable problems. For Cartesian lattice problems with possible grid compression, both D2Q9 and D2Q21 lattices give accurate results converging to the exact solutions for various values of \(\frac{1}{2}<\tau\leq 1\). For truly curvilinear cases considered we have presented evidence that the 4th-order isotropy of D2Q9 is insufficient and the D2Q21 lattice with 6th-order isotropy is needed for convergence to the exact solution. 

This illustration of the importance of moments' isotropy is very important. As it was stated in \cite{HChen}, a certain set of moment isotropy constraints and normalization conditions must be satisfied in order to correctly recover the full Navier-Stokes equations \cite{CGO,Shan,ChenZhang,ChenShan,Molvig}. In wide practical use are the small stencil length LBM lattices, such as D2Q9, largely due to their simplicity of implementation, specifically for the boundary conditions. However, as we discuss in Appendix A, D2Q9 lattice satisfies the requirements of moments' isotropy only to the 4-th order. Only the D2Q21 lattice satisfies those conditions to the 6th-order. 

The convergence to exact analytical solutions with the increase in resolution for various values of \(\tau\) we have obtained is quite good for all cases considered using the D2Q21 lattice. However, the simpler D2Q9 lattice can still be used only for Cartesian geometry cases with 1-dimensional grid compression. The curvilinear LBM method \cite{HChen},  although somewhat slower than the standard LBM method due to its higher mathematical complexity, still keeps all the advantages of the classical LBM method, such as intrinsic parallelism, applicability to complex physics cases (such as multi-phase flows, etc.), and no numerical diffusion at the advection stage. 

In order to perform our studies we have developed the generalized LBM boundary conditions of periodicity, no-slip, and moving wall types for the fully curvilinear geometry. We have additionally proposed a "no flow" adjustment procedure which helps to compensate for the effects of analytical finite-difference approximation for generalized basis vectors used in \cite{HChen}. 

We find our results to be very promising because, in our view, they open the much-needed way to expand the LBM method advantages to curvilinear geometries. The inability of the LBM method to adequately treat truly curvilinear cases and adaptive grids has been perceived as a weakness of LBM methods as compared to the classical finite-difference methods. In particular, such problems as adaptive grid compression into the boundary layer for high Reynolds number flows can possibly be addressed. 

Much work remains to be done, such as studies of 3-dimensional flows, extension to turbulent flow, extensions to multi-phase and multi-component flows, development of higher-order LBM boundary conditions for curvilinear cases for lattices with wider stencils such as D2Q21, and many others. 

Authors Alexei Chekhlov, Ilya Staroselsky, Raoyang Zhang, and Hudong Chen are employed by Dassault Systemes. The authors declare that the research was conducted in the absence of any commercial or financial relationships that could be construed as a potential conflict of interest. 

\begin{appendices}

\section{D2Q9 and D2Q21 Lattices}

\begin{figure}[H]
\centering
\caption{Component vectors}
\label{Lattices}
\begin{subfigure}[H]{0.6\textwidth}
\centering
\subcaption{D2Q9 lattice}
\label{D2Q9Lattice}
\includegraphics[width=\textwidth]{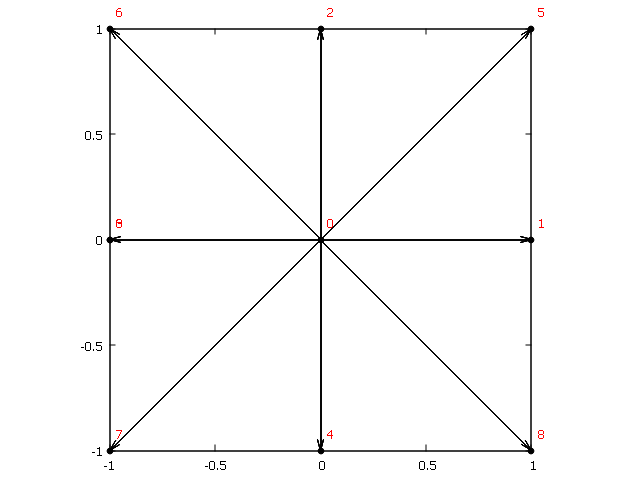}
\end{subfigure}
\begin{subfigure}[H]{0.6\textwidth}
\centering
\subcaption{D2Q21 lattice}
\label{D2Q21Lattice}
\includegraphics[width=\textwidth]{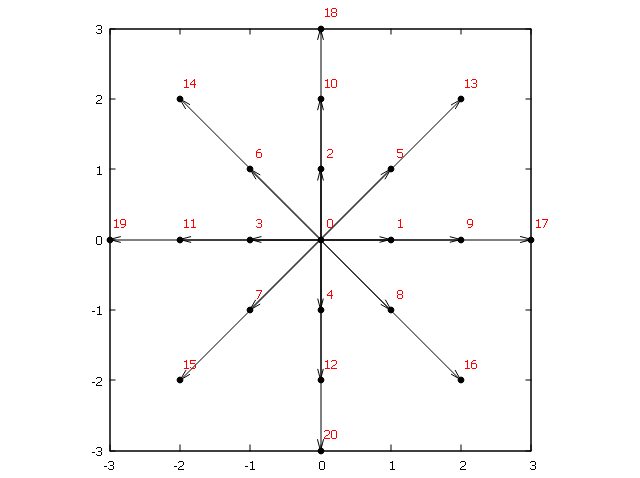}
\end{subfigure}
\end{figure}

The D2Q9 lattice shown in \Cref{D2Q9Lattice}
is defined by \(T_0=\frac{1}{3}\) and the following values of lattice component vectors and their weights shown in Tbl. \ref{D2Q9 lattice}: 

\begin{table}[H]
\centering
\scalebox{0.93}{
\begin{tabular}{|c|c|}
\hline
\(\alpha\) & \(w_{\alpha}\) \\
\hline
\(0\) & \(\frac{4}{9}\) \\
\(1\) & \(\frac{1}{9}\) \\
\(2\) & \(\frac{1}{9}\) \\
\(3\) & \(\frac{1}{9}\) \\
\(4\) & \(\frac{1}{9}\) \\
\(5\) & \(\frac{1}{36}\) \\
\(6\) & \(\frac{1}{36}\) \\
\(7\) & \(\frac{1}{36}\) \\
\(8\) & \(\frac{1}{36}\) \\
\hline
\end{tabular}}
\caption{D2Q9 lattice weights}
\label{D2Q9 lattice}
\end{table}

The D2Q21 lattice shown in \Cref{D2Q21Lattice} is defined by \(T_0=\frac{2}{3}\) and the following values of lattice component vectors and their weights shown in Tbl. \ref{D2Q21 lattice}: 
\begin{table}[H]
\centering
\scalebox{0.93}{
\begin{tabular}{|c|c|}
\hline
\(\alpha\) & \(w_{\alpha}\) \\
\hline
\(0\) & \(\frac{91}{324}\) \\
\(1\) & \(\frac{1}{12}\) \\
\(2\) & \(\frac{1}{12}\) \\
\(3\) & \(\frac{1}{12}\) \\
\(4\) & \(\frac{1}{12}\) \\
\(5\) & \(\frac{2}{27}\) \\
\(6\) & \(\frac{2}{27}\) \\
\(7\) & \(\frac{2}{27}\) \\
\(8\) & \(\frac{2}{27}\) \\
\(9\) & \(\frac{7}{360}\) \\
\(10\) & \(\frac{7}{360}\) \\
\(11\) & \(\frac{7}{360}\) \\
\(12\) & \(\frac{7}{360}\) \\
\(13\) & \(\frac{7}{432}\) \\
\(14\) & \(\frac{7}{432}\) \\
\(15\) & \(\frac{7}{432}\) \\
\(16\) & \(\frac{7}{432}\) \\
\(17\) & \(\frac{7}{1620}\) \\
\(18\) & \(\frac{7}{1620}\) \\
\(19\) & \(\frac{7}{1620}\) \\
\(20\) & \(\frac{7}{1620}\) \\
\hline
\end{tabular}}
\caption{D2Q21 lattice weights}
\label{D2Q21 lattice}
\end{table}

The set of even-order moment isotropy conditions up to the 6th order in \(\vec{c}\) are (all odd-order moments must be equal to \(0\)): 
\begin{equation}
\label{IsotropyConditions}
\begin{split}
\sum_{\alpha}w_{\alpha}=&1; \\
\sum_{\alpha}w_{\alpha}c_{\alpha}^ic_{\alpha}^j=&T_0\delta^{ij}\equiv T_0\triangle^{(2),ij}; \\
\sum_{\alpha}w_{\alpha}c_{\alpha}^ic_{\alpha}^jc_{\alpha}^kc_{\alpha}^l=&T_0^2\bigg[\delta^{ij}\delta^{kl}+\delta^{ik}\delta^{jl}+\delta^{il}\delta^{jk}\bigg]\equiv T_0^2\triangle^{(4),ijkl}; \\
\sum_{\alpha}w_{\alpha}c_{\alpha}^ic_{\alpha}^jc_{\alpha}^kc_{\alpha}^lc_{\alpha}^mc_{\alpha}^n=&T_0^3\bigg[\delta^{ij}\triangle^{(4),klmn}+\delta^{ik}\triangle^{(4),jlmn}+\delta^{il}\triangle^{(4),jkln}+\delta^{in}\triangle^{(4),jklm}\bigg]\equiv \\
\equiv &T_0^3\triangle^{(6),ijklmn}. 
\end{split}
\end{equation}

As it was shown in \cite{CGO,Shan}, only the D2Q21 lattice satisfies all the isotropy conditions above up to the 6th order. The D2Q9 lattice only satisfies them up to the 4th order. 

\section{Lattice Boundary Conditions}
Note that, as we have implemented both D2Q9 and D2Q21 lattices, we will write out explicitly the boundary conditions for naturally curvilinear and potentially radially non-equidistant lattices used only for the D2Q21 lattice below. But one can get the boundary conditions for the D2Q9 lattice by simply omitting all equations for the densities \(f_{\alpha}\) with indices \(\alpha\geq 9\). The boundary conditions below are formally written for both circular problems considered, but they are exactly the same for planar problems with the obvious change \(N_r\rightarrow N_x\). 

Also note that all our exact flow solutions are azimuth-independent and therefore the only dependence of the densities is upon the radial index \(i\), and all the \(j\) indices being the same for all formulas are omitted for brevity. The lattice is half-spaced, with the left boundary located at \(i=\frac{1}{2}\), the right boundary located at \(i=N_r+\frac{1}{2}\), the upper boundary located at \(j=N_{\theta}+\frac{1}{2}\), and the lower boundary located at \(j=\frac{1}{2}\). The internal lattice locations indices, therefore, are: \(1\leq i\leq N_r\), \(1\leq j\leq N_{\theta}\), and \(f_{\alpha}\) denotes post-advection distribution, and \(f_{\alpha}'\) denotes post-collision distribution. Note that as opposed to the standard LBM, this post-collision distribution includes the curvilinear correction in addition to the classical part, as described by the right-hand side of Eqs.  \mref{lbe,adv} above. As in \cite{HChen}, \(J\left(\vec{q}\right)\) denotes the area of the lattice cell centered at \(\vec{x}\left(\vec{q}\right)\). 

\subsection{Periodicity}

In the vertical, or azimuthal direction we used the periodicity condition, which can be implemented as follows. For such LBM velocity vectors \(\vec{c}_{\alpha}\) for which \(c_{\alpha}^2\neq 0\), or for \(\alpha\neq 1,8,9,11,17,19\),  we have the following general expression: 
\begin{equation}
N_{\alpha}\left(\vec{x},t+1\right)=N'_{\alpha}\left(\vec{x}-\vec{c}_{\alpha}, t\right). 
\end{equation}

\subsection{No-Slip}

In the horizontal, or radial direction we have used the no-slip boundary conditions, which were implemented as bounce-back with a modification for the variable lattice cell area. 

For such LBM velocity vectors \(\vec{c}_{\alpha}\) for which \(\left(c_{\alpha}\cdot\vec{n}\right)>0\), where \(\vec{n}\) is a surface normal pointing into the fluid, or for \(\alpha\neq 2,4,10,12,18,20\),  we have the following general expression: 
\begin{equation}
N_{\alpha}\left(\vec{x},t+1\right)=N'_{\alpha^*}\left(\vec{x}-\vec{c}_{\alpha}, t\right), 
\end{equation}
where the lattice nodes \(\vec{x}-\vec{c}_{\alpha}\) were extended through the boundaries using symmetry condition and the index \(\alpha^*\) is defined by the bounce-back conditions listed in \Cref{IndicesMapping}, \Cref{BounceBack}. 
\begin{table}[H]
\centering
\caption{Indices mapping}
\label{IndicesMapping}
\begin{subtable}{0.6\hsize}
\centering
\subcaption{Bounce-back}
\label{BounceBack}
\scalebox{0.93}{
\begin{tabular}{|c|c|}
\hline
\(\alpha\) & \(\alpha^*\) \\
\hline
\(1\) & \(3\) \\
\(3\) & \(1\) \\
\(5\) & \(7\) \\
\(6\) & \(8\) \\
\(7\) & \(5\) \\
\(8\) & \(6\) \\
\(9\) & \(11\) \\
\(11\) & \(9\) \\
\(13\) & \(15\) \\
\(14\) & \(16\) \\
\(15\) & \(13\) \\
\(16\) & \(14\) \\
\(17\) & \(19\) \\
\(19\) & \(17\) \\
\hline
\end{tabular}}
\end{subtable}
\begin{subtable}{0.6\hsize}
\centering
\subcaption{Specular Reflection}
\label{SpecularReflection}
\scalebox{0.93}{
\begin{tabular}{|c|c|}
\hline
\(\alpha\) & \(\hat{\alpha}\) \\
\hline
\(1\) & \(3\) \\
\(3\) & \(1\) \\
\(5\) & \(6\) \\
\(6\) & \(5\) \\
\(7\) & \(8\) \\
\(8\) & \(7\) \\
\(9\) & \(11\) \\
\(11\) & \(9\) \\
\(13\) & \(14\) \\
\(14\) & \(13\) \\
\(15\) & \(16\) \\
\(16\) & \(15\) \\
\(17\) & \(19\) \\
\(19\) & \(17\) \\
\hline
\end{tabular}}
\end{subtable}
\end{table}

\subsection{Moving Wall}

The moving wall boundary condition was implemented in the horizontal or radial direction as specular reflection with moving wall velocity correction and the modification to reflect the variable lattice cell area. 

For such LBM velocity vectors \(\vec{c}_{\alpha}\) for which \(\left(c_{\alpha}\cdot\vec{n}\right)>0\), where \(\vec{n}\) is a surface normal pointing into the fluid, or for \(\alpha\neq 2,4,10,12,18,20\),  we have the following general expression: 
\begin{equation}
N_{\alpha}\left(\vec{x},t+1\right)=N'_{\hat{\alpha}}\left(\vec{x}-\vec{c}_{\hat{\alpha}}, t\right)+\frac{2}{T_0}\rho\left(\vec{x}-\vec{c}_{\hat{\alpha}},t\right)c_{\hat{\alpha}}^j\left( U_0^j-\tilde{U}^j\left(\vec{x}-\vec{c}_{\hat{\alpha}},t\right)\right), 
\label{SpecularReflectionBCs}
\end{equation}
where, as everywhere in \cite{HChen}, the summation over \(j\) is assumed and where the lattice nodes \(\vec{x}-\vec{c}_{\alpha}\) were extended through the boundaries using symmetry condition and the index \(\hat{\alpha}\) is defined by the specular reflection boundary conditions listed in \Cref{IndicesMapping}, \Cref{SpecularReflection}. 

Note that in the Eq. \eqref{SpecularReflectionBCs} we have neglected the first term corresponding to \(j=1\) in our calculations since only the azimuthal flow is assumed. 

\section{Linearly Contracting Grid}

Within an interval \(x\in\left[0,l\right]\) let us build a non-equidistant lattice that will be used in most of the considered exact solutions. This lattice will be contracting from the middle of the domain \(x=\frac{l}{2}\) towards the boundaries \(x=0\) and \(x=l\) in a linear fashion. 

First, let us formulate such boundaries-aligned lattice, \(x_i'\). For that, let us assume that the lattice size \(N\) is even, \(N=2K\), and denote the variable lattice step size as \(\triangle_i'=x_i'-x_{i-1}'\). 

In the left part of the domain \(0\leq x\leq \frac{l}{2}\) we seek the lattice locations in the following form: 
\[
\triangle_i'=a\overline{\triangle}i+b,
\]
where we have denoted the average lattice step as \(\overline{\triangle}=\frac{l}{N}\). 
The general solution for lattice locations that satisfies boundary conditions \( x_0'=0\) and \(x_{\frac{N}{2}}'=\frac{l}{2}\) for \( 0\leq i\leq\frac{N}{2}\) is: 
\[
x_i'=\overline{\triangle}i\bigg\{1+\frac{a}{2}\left(i-\frac{N}{2}\right)\bigg\}, 
\]
and for the lattice steps is: 
\[
\triangle_i'=\overline{\triangle}\bigg\{1+a\left(i-\frac{N}{4}-\frac{1}{2}\right)\bigg\}.
\]
As intended, for \(a>0\) the lattice steps linearly vary from the smallest at the left boundary \(x_0'=0\): 
\[
\triangle_1'=\overline{\triangle}\bigg\{1-\frac{a}{2}\left(\frac{N}{2}-1\right)\bigg\}
\]
to the largest in the middle \(x_{\frac{N}{1}}=\frac{l}{2}\): 
\[
\triangle_{\frac{N}{2}}'=\overline{\triangle}\bigg\{1+\frac{a}{2}\left(\frac{N}{2}-1\right)\bigg\}.
\]
Requiring the smallest lattice step to be positive leads to the following condition on the contraction parameter \(a\): 
\[
0\leq a\leq a_{max}(N)=\frac{2}{\frac{N}{2}-1}.
\]
Similarly, in the right part of the lattice domain \(\frac{N}{2}+1\leq i \leq N\) we can get for the lattice coordinates: 
\begin{equation}
\label{LinearLatticePositions1}
x_i'=\overline{\triangle}\bigg\{\frac{N}{2}+\left(i-\frac{N}{2}\right)\left(1+\frac{a}{2}\left( N-i \right)\right)\bigg\}, 
\end{equation}
and for the lattice steps: 
\begin{equation}
\label{LinearLatticeSteps1}
\triangle_i'=\overline{\triangle}\bigg\{1-a\left(i-\frac{3N}{4}-\frac{1}{2}\right)\bigg\}. 
\end{equation}
Using this grid-aligned lattice, we can easily define the requirements for LBM half-spaced lattice as follows: 
\begin{equation}
\label{LinearLatticePositions2}
x_i=\frac{x_i'+x_{i-1}'}{2},
\end{equation}
and its steps: 
\begin{equation}
\label{LinearLatticeSteps2}
\triangle_i=\frac{x_i'-x_{i-2}'}{2}.
\end{equation}
Using these results we can introduce the Contraction Ratio (CR) as the ratio by how much the first step is different from the average or equidistant one: 
\begin{equation}
CR=\frac{\overline{\triangle}-\triangle_1'}{\overline{\triangle}}, 
\end{equation}
so that \(CR=0\) corresponds to the equidistant case and \(CR>0\) corresponds to steps contraction towards \(i=1\). Using this we can also introduce the Steps Ratio (SR), as: 
\begin{equation}
SR\equiv\frac{\triangle_1'}{\triangle_{\frac{N}{2}}'}=\frac{1-CR}{1+CR}. 
\end{equation}
An example of a rectangular grid with compression with \(CR=0.4\) along the \(x-\)axis and equidistant along the \(y-\)axis which was used in our planar problems is shown in Fig. \ref{GridContractionPlanarGeometry}. 

\section{Relationship Between General and Polar Basis Vectors}

First, let us specify the basis vectors defined in \cite{HChen} to the unit basis vectors in the polar coordinate system on the half-spaced lattice. 

The unit basis vectors in polar coordinates on a half-spaced lattice are: 
\begin{equation}
\begin{split}
\hat{r}=&\frac{\frac{\partial\vec{r}}{\partial r}}{\big|\frac{\partial\vec{r}}{\partial r}\big|}=
\begin{pmatrix}
\cos\bigg[\left(j-\frac{1}{2}\right)\triangle_{\theta}\bigg] \\
\sin\bigg[\left(j-\frac{1}{2}\right)\triangle_{\theta}\bigg]
\end{pmatrix}; \\
\hat{\theta}=&\frac{\frac{\partial\vec{r}}{\partial\theta}}{\big|\frac{\partial\vec{r}}{\partial\theta}\big|}=
\begin{pmatrix}
-\sin\bigg[\left(j-\frac{1}{2}\right)\triangle_{\theta}\bigg] \\
\cos\bigg[\left(j-\frac{1}{2}\right)\triangle_{\theta}\bigg]
\end{pmatrix},
\end{split}
\end{equation}
where \(1\leq j\leq N_{\theta}\) and \(\triangle_{\theta}=\frac{2\pi}{N_{\theta}}\). 

The tangent basis vectors for current polar geometry, as defined in \cite{HChen}, are: 
\begin{equation}
\begin{split}
\vec{g}_1=&\frac{r_{i+1}-r_{i-1}}{2r_i\overline{\triangle}_r}\vec{r}=\frac{r_{i+1}-r_{i-1}}{2\overline{\triangle}_r}\hat{r}; \\
\vec{g}_2=&\frac{\sin\triangle_{\theta}}{\overline{\triangle}_r}A_{\frac{\pi}{2}}\vec{r}=\frac{\sin\triangle_{\theta}}{\overline{\triangle}_r}r_i\hat{\theta},
\end{split}
\end{equation}
where matrix \(A_{\phi}\) is a matrix of counter-clockwise rotation by an angle \(\phi\): 
\begin{equation}
A_{\phi}=
\begin{pmatrix}
\cos\phi & -\sin\phi \\
\sin\phi & \cos\phi
\end{pmatrix}.
\end{equation}
Therefore, for any vector component representation  \(\vec{u}=u_r\hat{r}+u_{\theta}\hat{\theta}=U^1\vec{g}_1+U^2\vec{g}_2\) we have the following correspondence between its general curvilinear expansion in \(\vec{g}_1,\vec{g}_2\) and in polar \(\hat{r},\hat{\theta}\) basis vectors: 
\begin{equation}
u_r=U^1\frac{r_{i+1}-r_{i-1}}{2\overline{\triangle}_r}, \mbox{ and } u_{\theta}=U^2\frac{\sin\triangle_{\theta}}{\overline{\triangle}_r}r_i.
\end{equation}
\end{appendices}

\printbibliography

\end{document}